\documentclass[preprint, superscriptaddress]{revtex4}
\usepackage{graphicx} % include figure files 
\usepackage{bm} % bold math 
\usepackage{multirow} 
\usepackage{subfigure}
\usepackage{amsfonts}
\usepackage{color}
\begin{document}
 
\title{Effects of nano-void density, size, and spatial population on thermal conductivity: a case study of GaN crystal}
 
\author{X. W. Zhou}
\email[]{X. W. Zhou: xzhou@sandia.gov}
\affiliation{Mechanics of Materials Department, Sandia National Laboratories, Livermore, California 94550, USA}

\author{R. E. Jones}
\affiliation{Mechanics of Materials Department, Sandia National Laboratories, Livermore, California 94550, USA}

\date{\today}
 
\begin{abstract}

The thermal conductivity of a crystal is sensitive to the presence of surfaces and nanoscale defects. While this opens tremendous opportunities to tailor thermal conductivity, a true ``phonon engineering'' of nanocrystals for a specific electronic or thermoelectric application can only be achieved when the dependence of thermal conductivity on the defect density, size, and spatial population is understood and quantified. Unfortunately, experimental studies of effects of nanoscale defects are quite challenging. While molecular dynamics simulations are effective in calculating thermal conductivity, the defect density range that can be explored with feasible computing resources is unrealistically high. As a result, previous work has not generated a fully detailed understanding of the dependence of thermal conductivity on nanoscale defects. Using GaN as an example, we have combined physically-motivated analytical model and highly-converged large scale molecular dynamics simulations to study effects of defects on thermal conductivity. An analytical expression for thermal conductivity as a function of void density, size, and population has been derived and corroborated with the model, simulations, and experiments. 

\end{abstract}
 
% inserts suggested PACS numbers in braces on next line \pacs{66.70,-f, 02.70.Ns, 05.60.Cd, 44.10.+i}

% insert suggested keywords-APS authors don't need to do this
%\keywords{}

\maketitle

% body of paper here-Use proper section commands 
% References should be done using the\cite, \ref, and \label commands

\section{Introduction}

Ability to tailor thermal transport properties of materials has become increasingly important in modern society. In one example, reducing thermal conductivity of thermoelectric materials makes it practical to generate electricity from waste heat \cite{HLGCDUHPK2004,MSS1997}. In another example, increasing thermal conductivity of the semiconductor materials enables an increase in the device density by improving heat dissipation \cite{PN2006,S2006}. It is well known that thermal conductivities of materials are extremely sensitive to defect populations \cite{NIBBL1991,PM2009,AMNST1999,YMMCDU2003,ZUJC2005,PCCRDLH2006,WPXQWQF2006}. Control of defects therefore provides a powerful means to minimize thermal conductivities for thermoelectric applications. The presence of defects, however, is a concern to the electronic devices especially in the nano-scale. This is because defects not only affect electronic properties, but also cause phonon scattering that reduces thermal conductivities. This can lead to abnormal heating in local regions around defects. If the heated region is the bottle-neck of a nanostructure, for example, a cross-section of a nanowire, catastrophic failure may be triggered \cite{WJHWLT2009}. A detailed understanding of thermal conductivity as a function of defects is thus essential.

Given experimental evidence of significant defect incorporation \cite{RM2005,ALBLTT2009,AWT2009,FVSLME2008,TWLA2008}, studies of defects in GaN nanostructures are of particular interest. Specifically, recent experiments \cite{LW2010} suggest that the spatial distribution of defects in GaN nanowires is not uniform, and vacancies can combine to form nano-scale voids near surfaces. Understanding thermal conductivity as a function of void population using GaN as an example is therefore technologically beneficial since GaN is an important semiconductor that can be readily integrated with Si and is ideally suited for optoelectronic applications \cite{SSAMD1993}. 

Various experiments have been performed to quantify thermal conductivity as a function of porosity of porous materials \cite{SC2004,BBS1997,R1996,BBMMR2005,R1975}. However, experimental studies of nanoscale voids are much more challenging and a detailed knowledge of void effects has not been developed in experiments. Extensive theoretical work has also been performed to develop relationships between thermal conductivity and point defects \cite{CV1960,K1955,A1963,S1957,HK1993,YMMCDU2003,AMNST1999,LPY1998}. Most of these studies \cite{CV1960,K1955,A1963,S1957,HK1993,YMMCDU2003,AMNST1999} used an analytical approach initially developed by Callaway and Vonbaeyer \cite{CV1960}, Klemens \cite{K1955}, Abeles \cite{A1963}, and Slack \cite{S1957} et al. In this approach, a point defect phonon scattering relaxation time is expressed as a function of the atomic masses, and the strain caused by the defects. This relaxation time is then used in Debye model to calculate thermal conductivity. While this approach has been successfully applied to explore thermal conductivity as a function of defect density, the defect scattering parameter in the model is often treated as a fitting parameter requiring experimental data to determine. Also, this model has not been used to distinguish different types of defects such as vacancies and voids (with different sizes). In particular, we note that the model was mainly used for point defects but not for bigger defects such as voids.

By using a discrete lattice directly in a computational system where defects of different sizes and shapes can be precisely replicated on an atomistic scale, molecular dynamics (MD) enables the effects of voids on thermal conductivity to be predicted from a known interatomic potential without additional assumptions. MD therefore provides a powerful means to study thermal transport \cite{LPY1998,VC2000,SPK2002,ZAJGS2009,CCG2000,JJ1999,CFP2011,FP2011,SGOS2011,TOT2010,CP2011}. In particular, MD has been used to study the effects of vacancy, interstitial, and antisite defects on thermal conductivity \cite{LPY1998,SGOS2011,TOT2010,CP2011}, and to determine the scattering of phonon wave packets by point defects \cite{YWSKCP2008}. Using relatively small systems imposed by the computational cost, these previous MD studies were limited to very high defect density (site fraction 0.5\% \cite{LPY1998}, $\geq$ 0.0125\% \cite{SGOS2011}, $\geq$ 0.39\% \cite{TOT2010}, $\geq$ 0.075\% \cite{CP2011}, and $\geq$ 0.016\% \cite{YWSKCP2008}) even when a single defect is simulated. In contrast, a very high experimental vacancy concentration of $10^{15}~cm^{-3}$ only corresponds to a site fraction of $1.15\times10^{-6}$\%. It is unclear how the MD results scale with the experimental data obtained at defect densities orders of magnitude smaller.

\begin{figure}
\includegraphics[width=4in]{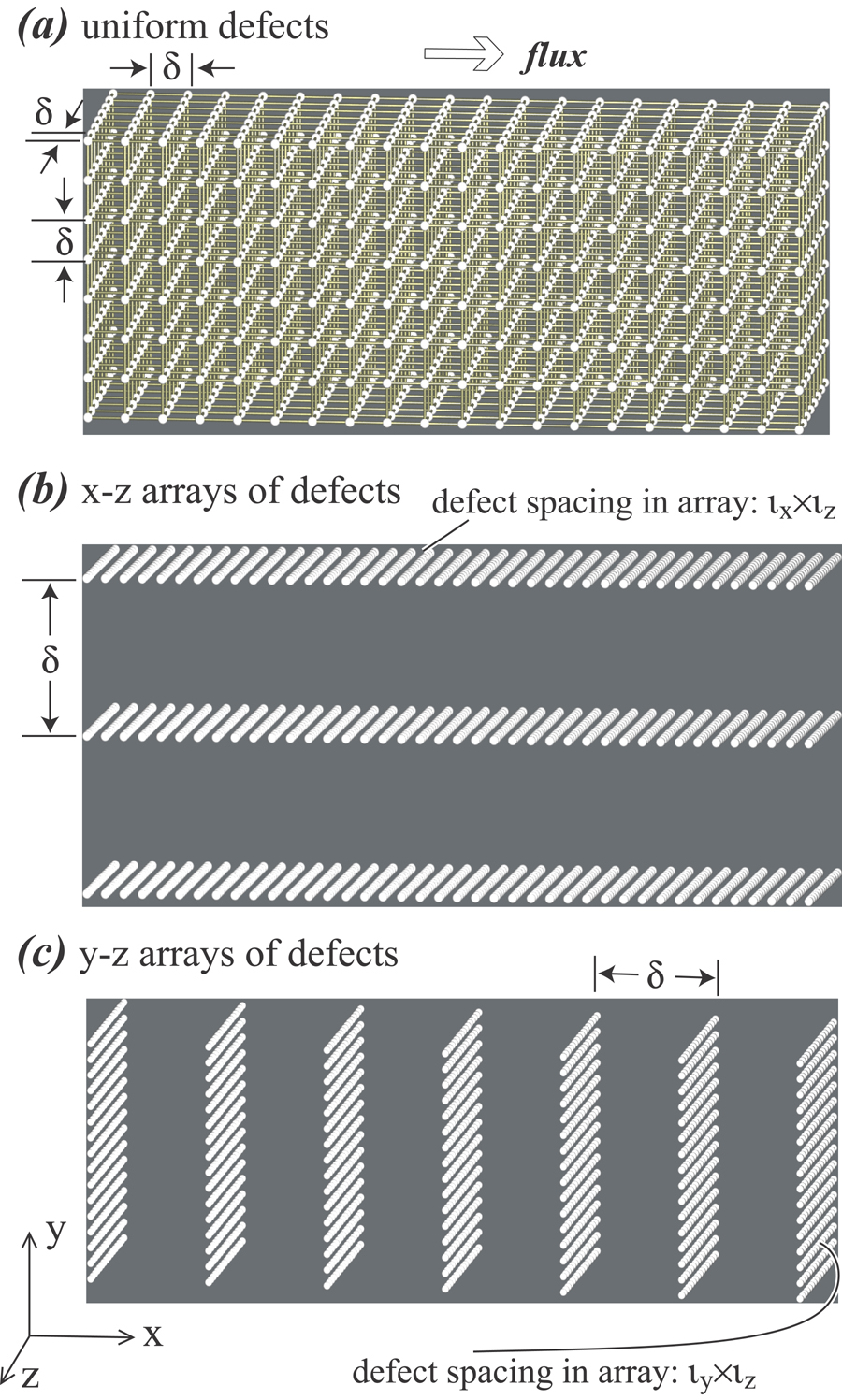}
\caption{Three void distributions: (a) a bulk configuration with a uniform density; (b) a configuration with arrays of voids on the x-z plane; and (c) a configuration with arrays of voids on the y-z plane.
\label{defect arrays}}
\end{figure}

Recently, we have developed and verified a scaling law \cite{ZJA2010a} that can extend the size-dependent MD conductivity estimates to the dimensions of realizable devices. In particular, analytical expressions of thermal conductivity as a function of film thickness or wire radius are derived \cite{ZJA2010b}. Such a scaling law provides an ideal means to evaluate the scaling of defect density from MD to experimental conditions (a discussion of the validity of the scaling law is given in Appendix \ref{length scaling}). Here the scaling law of defects can be explored using three defect distributions as shown in Fig. \ref{defect arrays}. In Fig. \ref{defect arrays}(a), the defects are uniformly distributed in all three directions with equal spacing $\delta$. In Fig. \ref{defect arrays}(b), defects are closely spaced in the $x-$ and $z-$ directions with spacings of $\iota_x$ and $\iota_z$ respectively but the $x-z$ defect arrays have a large spacing $\delta$ in the $y-$ direction. In Fig. \ref{defect arrays}(c), defects are closely spaced in the $y-$ and $z-$ directions with spacings of $\iota_y$ and $\iota_z$ respectively but the $y-z$ defect arrays have a large spacing $\delta$ in the $x-$ direction. Increasing $\delta$ can effectively reduce the void site fraction. Despite different distributions, the mechanism and the functional dependence of thermal conductivity on $\delta$ is similar in all three cases if the spacings in each defect array (i.e., $\iota_x$, $\iota_y$, $\iota_z$) remain constant. Compared with Fig. \ref{defect arrays}(a), smaller systems can be used in Figs. \ref{defect arrays}(b) and \ref{defect arrays}(c) to capture the same defect density. Hence, configurations shown in Figs. \ref{defect arrays}(b) and \ref{defect arrays}(c) are more suitable for the computationally extremely expensive MD simulations. 

The present work addresses the scaling of defect effects on thermal conductivity in the realistic defect density range using three integrated approaches: (a) develop a physically-motivated model and use it to derive an analytical expression of thermal conductivity as a function of void density, size, and spatial population; (b) verify the analytical model using massively parallel MD simulations in a case study that studies the effects of voids on the thermal conductivity of GaN wurtzite crystal; and (c) corroborate the analytical model and MD simulations using the extensive experimental thermal conductivity data of porous materials. The significance of the recent experimental observation that the nano-voids in some GaN nanowires appear mostly near the surfaces rather than in the bulk \cite{LW2010} will also be evaluated.

\section{Molecular Dynamics Simulation Methodology}

Molecular dynamics simulations are used to both develop and verify the analytical model of defects. In particular, the ``direct method'' \cite{ZJA2010a,ZJA2010b,ZAJGS2009} was used to calculate the $[0001]$ thermal conductivity / resistivity of a defective wurtzite GaN crystal at a temperature of T = 300 K. Here the temperature was defined as $T = 2e_k/(3k)$, where $e_k$ is average kinetic energy per atom and $k$ is Boltzmann constant. We caution that the 300 K temperature is at the lower bound of the estimated Debye temperature range of 350 - 600 K \cite{S1973,MAG1998,DPWJP2006}, and we did not perform quantum correction on temperature as it has been found to be of dubious benefit \cite{TMA2009}. However, the use of the low temperature is satisfactory in the present work because our purpose is to explore the scaling of thermal conductivity with defect density rather than evaluate the accuracy of the MD method on thermal conductivity calculations at low temperatures. One particular reason to choose the relatively low temperature is to reduce intrinsic noise of the data \cite{ZJA2010a,ZJA2010b,ZAJGS2009} so that the scaling relationship can be more reliably identified. Previous MD work \cite{ZAJGS2009,ZJA2010a,ZJA2010b} has employed the Stillinger-Weber (SW) potential parameterized by B\'er\'e and Serra \cite{BS2002,BS2006} to calculate the thermal conductivity of GaN bulk and nanostructured crystals. To compare with the previous results, we use the same potential in the present study. This potential gives reasonable prediction on dispersion relations, vibrational density of states (DOS), and heat capacity for the bulk GaN system \cite{ZAJGS2009}. To ensure accuracy of our MD thermal conductivity data, all of our calculations are based upon highly converged temperature profiles averaged over at least 50 ns, which is significantly longer than in similar studies. 

\begin{figure}
\includegraphics[width=6in]{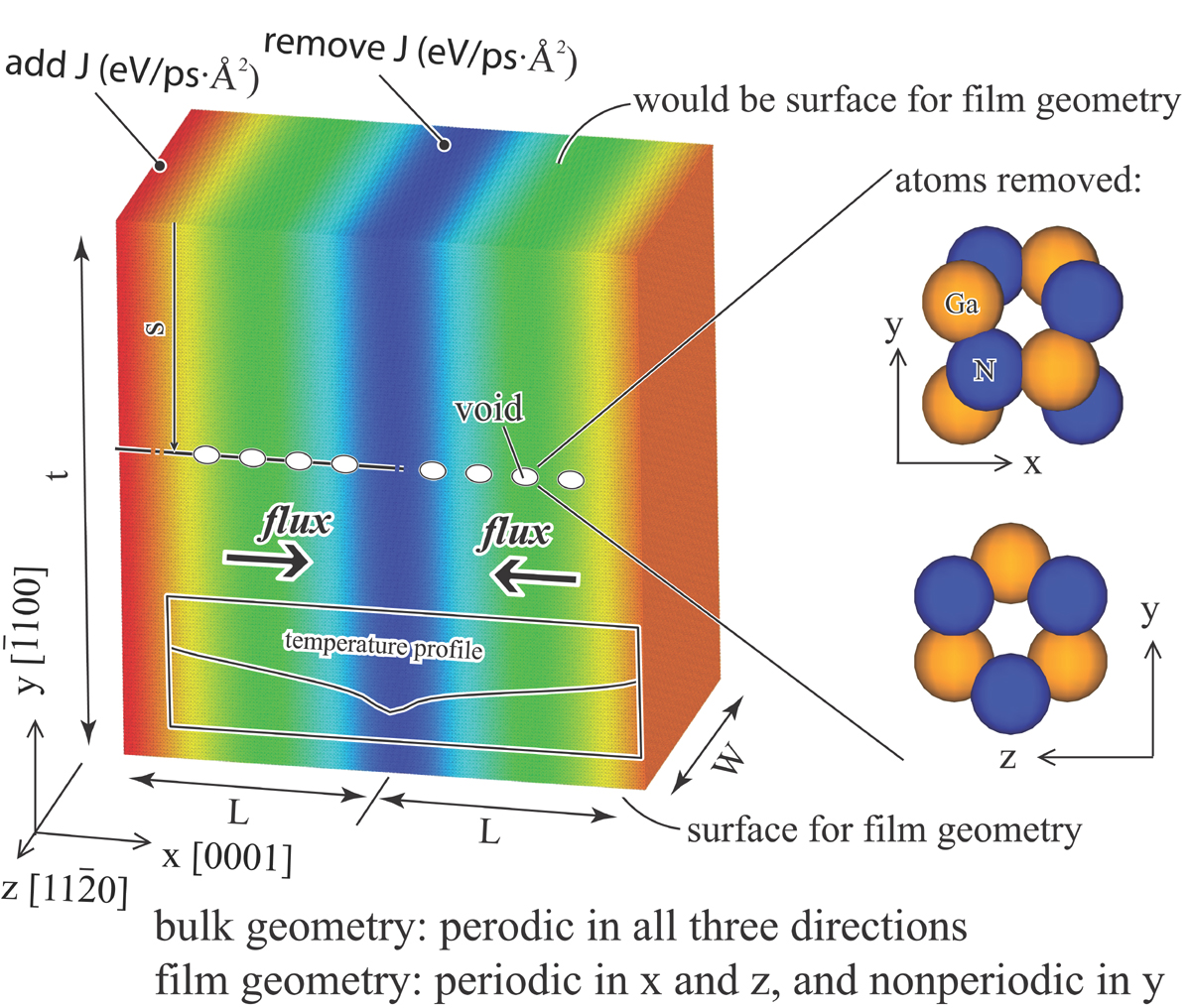}
\caption{Geometry of ``direct method'' MD simulations of thermal conductivity. 
\label{MD geometry}}
\end{figure}

The computational system used for the simulations is shown in Fig. \ref{MD geometry}. The hexagonal wurtzite GaN crystal is aligned so that the $x-$, $y-$, and $z-$ coordinates correspond respectively to $[0001]$, $[\bar1100]$, and $[11\bar20]$ directions. The computational system has a length $2L$ in the $x-$ direction, a thickness $t$ in the $y-$ direction, and a width $W$ in the $z-$ direction. According to the lattice constants of GaN and the chosen geometry, the smallest orthogonal cell has a dimension of $a_1 = c =$ 5.2000 \AA, $a_2 = 2\cdot a \cdot cos\left(\pi/6\right) =$ 5.5252 \AA, and $a_3 = a =$ 3.1900 \AA \ in the $x-$, $y-$, and $z-$ directions \cite{ZAJGS2009}. For convenience, the system dimensions are represented by the number of cells $n_1$, $n_2$, and $n_3$ in the $x-$, $y-$, and $z-$ directions. A constant heat flux method \cite{IH1994,SP2001,JJ1999,SPK2002,YCSS2004,ZAJGS2009,ZJA2010a,ZJA2010b} is used to create a temperature gradient in the $x-$ direction as shown by the color scheme in Fig. \ref{MD geometry} (red and blue mean respectively the highest and the lowest temperatures). Void defects are created uniformly between heat source and sink by removing clusters of atoms. In particular, the void is equivalent to a volume of 12 atoms and a shape of hexagonal column along the $x-$ direction as shown in the blow-up of Fig. \ref{MD geometry}. Such hexagonal voids are terminated with low index surfaces and are stable during simulations.

Thermal transport (along $x-$ direction) in both bulk and film configurations is considered. As shown in Fig. \ref{MD geometry}, the bulk means that the system thickness in the $y-$ direction is infinite, and this is achieved by using a periodic boundary condition in the $y-$ direction. The film means that the system thickness in the $y-$ direction is finite, and this is achieved by using a free boundary condition in the $y-$ direction. The periodic boundary conditions are always used in the $x-$ and $z-$ directions for both bulk and film configurations. It should be noted, however, that the phonon mean free path in the $x-$ direction is limited by the finite distance between the heat source and sink despite the periodic boundary condition. 

\section{Analytical Model of the Thermal Conductivity of Material with Voids}

\begin{figure}
\includegraphics[width=4in]{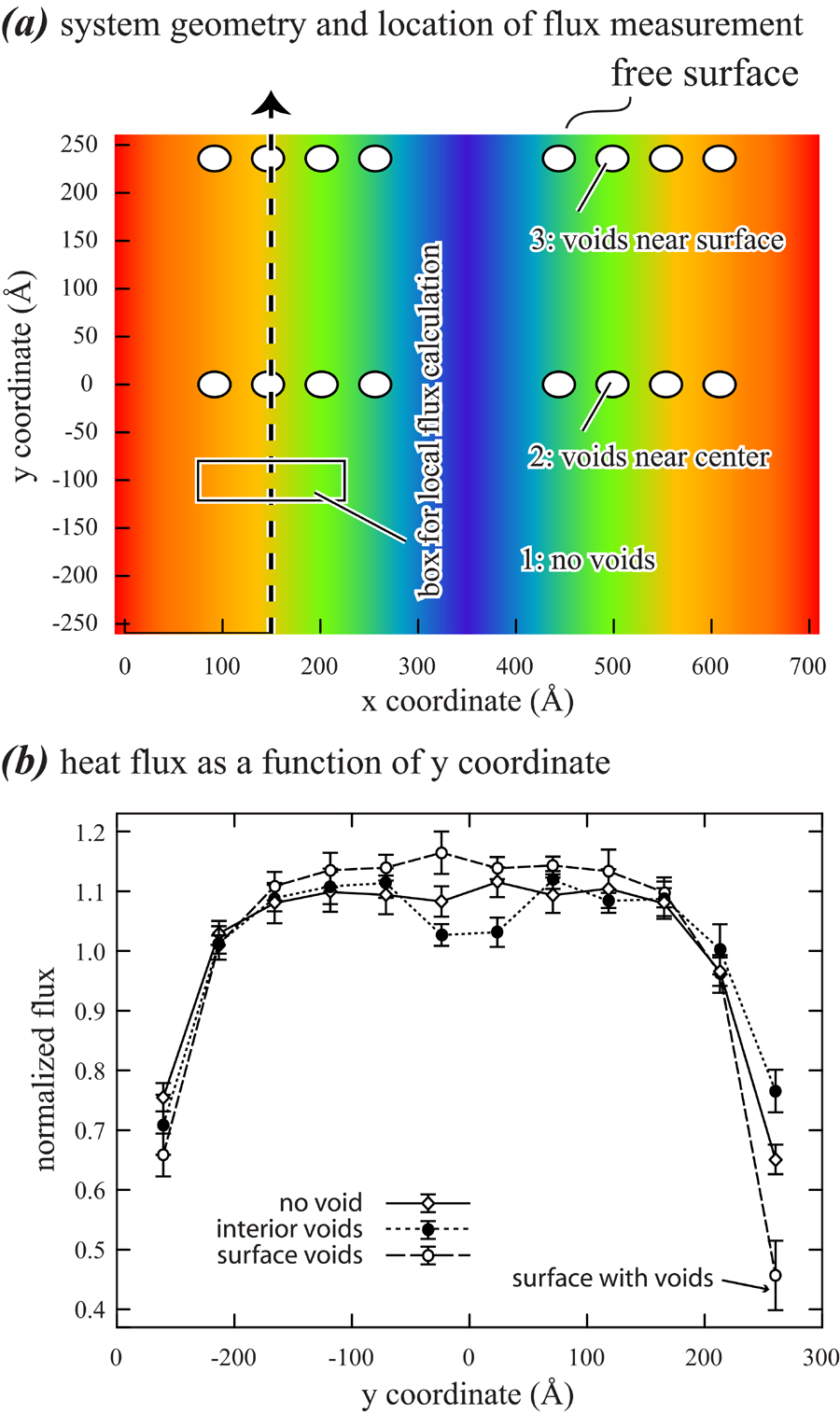}
\caption{Heat flux distributions in the cross-section of a thin film: (a) system geometry illustrating three void configurations (1: no voids; 2: voids near center; and 3: voids near one of the surfaces), box for calculating the local heat flux, and the $x-$ coordinate where the box is placed and moved in the $y-$ direction to calculate heat flux as a function of $y-$ coordinate; and (b) heat flux as a function of $y-$ coordinate for the three void configurations. The heat flux is normalized by the average flux value to emphasize the relative differences in flux across the film.
\label{flux_distribution}}
\end{figure}

To ensure that our analytical model of defects is physically well-motivated, MD simulations are first performed to examine the effect of defects on the heat flux distribution. A film configuration composed of around 590,000 atoms with dimensions of $n_1 = 136$ ($2L \approx 707 \AA$), $n_2 = 90$ ($t \approx 500 \AA$), and $n_3 = 6$ ($W \approx 19 \AA$) is used. Following the same methodology described previously \cite{ZJA2010b}, a local heat flux is calculated as a long time average of the atomic contributions to the overall heat flux \cite{SPK2002} for atoms in the rectangular box illustrated in Fig. \ref{flux_distribution}(a). The box has 150 \AA~ $x-$ dimension and about 41 \AA~ $y-$ dimension, and extends all the $z-$ dimension of computational cell. It is placed at an $x-$ coordinate of about 150 \AA. By moving the box along the $y-$ direction, heat flux is calculated as a function of y. Three simulations, corresponding respectively to: no defects, four voids equally spaced between the heat source and sink along the center plane of the film, and the same four voids along a plane near the $y-$ surface of the film (precisely 2 unit cells from the surface), are conducted as shown in Fig. \ref{flux_distribution}(a). The heat fluxes obtained from the three simulations are shown in Fig. \ref{flux_distribution}(b). 

Fig. \ref{flux_distribution}(b) indicates that for the sample with no defects, the heat flux inside the film remains relatively constant and the heat flux decreases only near the surfaces due to the surface scattering effect, as reported in previous work \cite{ZJA2010b}. From the graph, the scattering distance of the surface-affected regions can be estimated to be about 100 \AA. This is in agreement with previous molecular dynamics calculations of thermal conductivities of wires and films \cite{ZJA2010a,ZJA2010b} which indicated that at least for the GaN SW potential \cite{BS2002,BS2006}, the surface scattering affected heat flux is confined to a near-surface layer around 100 \AA~deep. Fig.~\ref{flux_distribution} also indicates that voids in the center of the film causes a reduction of heat flux due to the defect scattering effect. Apparently, the dimension of the affected region is also about 100 \AA~ (or less) for voids of this size. The heat flux in regions away from the voids and the surfaces is nearly constant and comparable to the values of the film without a defect. When the voids are moved to near the surface, the void and surface scattering zones are consolidated so that only a combined reduction of heat flux near the surface is observed. The difference is apparent in the reduction at the surface with the near-surface void which is slightly more significant than that at the other surface. These findings are the basis for the analytical model of the heat flow to be developed in the following.

\begin{figure}
\includegraphics[width=4in]{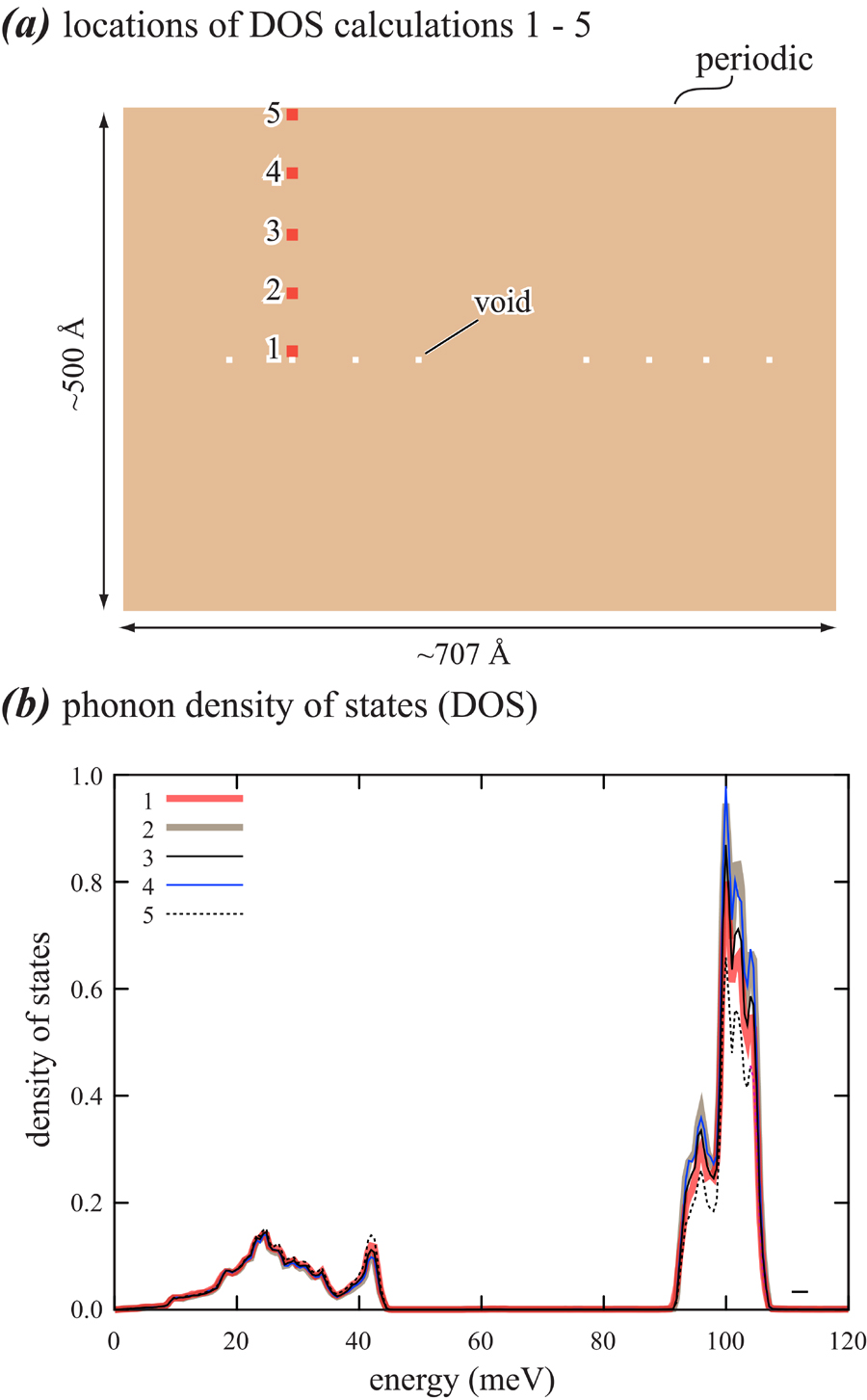}
\caption{The phonon density of states for a periodic system: (a) locations of 5 regions (red boxes) for calculations of density of states (from region 1 closest to the voids to region 5 furthest); and (b) the density of states calculated for the 5 locations shown in (a). Here the values of density of states have error bars of near the width of the variation of the acoustic modes (0-45 meV) and are omitted for clarity. The density of acoustic modes appears to be equivalent but the density of optical modes increases slightly near the void.
\label{void_dos}}
\end{figure}

Phonon density of states is also explored. The same system dimension as that in Fig. \ref{flux_distribution} is used with four voids equally spaced between heat source and sink along the center plane. Periodic boundary condition is used in the $y-$ direction to eliminate the surface effects so that the void effects can be more clearly revealed. Phonon density of states is calculated for five selected regions shown in Fig.~\ref{void_dos}(a) using the real scale (red boxes), where region 1 to region 5 has increasing distance from the voids. From the calculated density of states shown in Fig.~\ref{void_dos}(b), it is apparent that there is no significant change in the acoustic density of states across the system and there is a subtle increase in the optical density of states. The increase in optical states near the void is plausible. However, a full understanding of the influence of voids on individual phonons and the contributions of phonons to the observed flux behavior would require analysis of local mean free paths that is beyond the present scope \cite{HC2008}.

\subsection{Introduction of Model Concepts}

\begin{figure}
\includegraphics[width=6in]{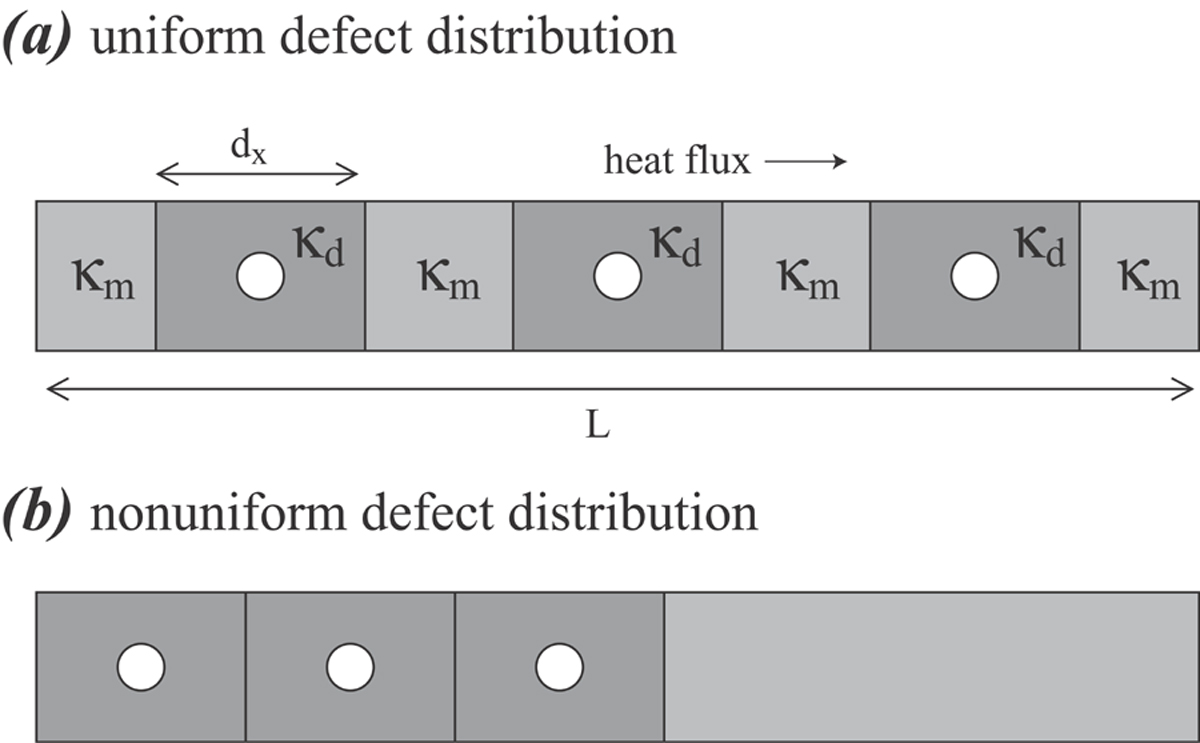}
\caption{Comparison of two defect distributions: (a) uniform and (b) nonuniform defect distributions.
\label{uniformity}}
\end{figure}

According to Fig. \ref{flux_distribution}, we assume that the mean effects of phonon scattering at surfaces and voids is confined to the 100 \AA~ ranges, at least for voids of comparable sizes to those simulated. Considering that the average spacing between point defects such as vacancies exceeds 1000 \AA~even when the material contains a high defect concentration of $10^{15}~cm^{-3}$, we can assume that defects are dispersed (i.e., isolated and independent). For convenience, we further assume that defects are uniformly distributed, which is roughly true for the most homogeneous experimental material, but is exact for simulations with periodic boundary conditions where defects are replicated uniformly in space. On the other hand, when the defect scattering effect is fully confined to a relatively small region, the overall thermal conductivity becomes independent on the particular distribution of sparse defects. This is examined in Fig. \ref{uniformity} using heat transport through a prism as an example. For sparse defects, we can embed each defect (circle) in a section (dark box) with a sufficiently large dimension of $d_x$. Because defect effects are confined to a relatively small region, we can assume that only the thermal conductivity inside the section changes to $\kappa_d$ whereas the conductivity outside the section remains to be the defect-free matrix value $\kappa_m$. It is plausible that, for the two cases where the three defects are either uniformly distributed over the prism length $L$, Fig. \ref{uniformity}(a), or concentrated on the left-hand side, Fig. \ref{uniformity}(b), the overall thermal resistivities (inverse of thermal conductivity) of the two prisms are the same, both equal to a length weighted average of resistivities of defective sectors and matrix \cite{ZJA2010a,ZJA2010b} as: $3d_x/L \cdot \kappa_d^{-1} + \left(L-3d_x\right)/L \cdot \kappa_m^{-1}$. Note that our model will be accurate when these assumptions are valid. The model may still capture well the scaling law even for materials with phonon mean free path longer than defect spacing. Model evaluation should be performed by comparing model predictions with either experiments or direct simulations (such as MD simulations as will be shown in the following for the GaN case).

\begin{figure}
\includegraphics[width=6in]{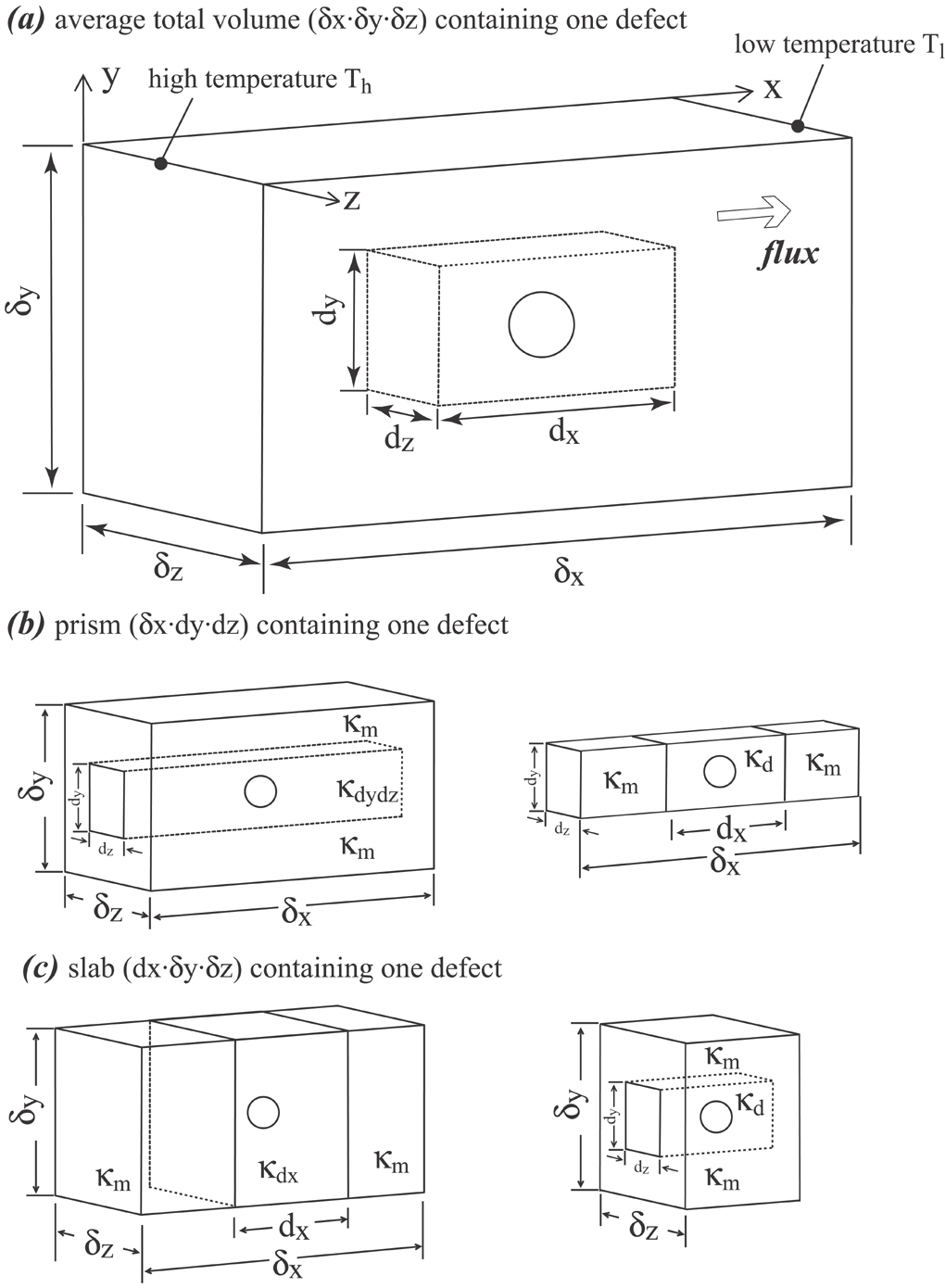}
\caption{Heat conduction through a volume $\delta_x \cdot \delta_y \cdot \delta_z$ of material containing one defect: (a) average total volume ($\delta x \cdot \delta y \cdot \delta z$) containing one defect; (b) prism ($\delta x \cdot dy \cdot dz$) containing one defect; and (c) slab ($dx \cdot \delta y \cdot \delta z$) containing one defect.
\label{model}}
\end{figure}

When the defect distribution is uniform, the material can be viewed as composed of equivalent small volumes $\delta V = \delta_x \cdot \delta_y \cdot \delta_z$ each containing a defect. Because these volumes are assumed to be identical and independent of each other, the overall thermal conductivity of the material equals the thermal conductivity of each individual volume. As a result, only one representative volume needs to be addressed. Here, we consider heat conduction in the x dimension of the volume $\delta V = \delta_x \cdot \delta_y \cdot \delta_z$ containing a defect in its center as marked by a spherical ball in Fig. \ref{model}(a). The presence of this defect changes the thermal conductivity of its surroundings defined by a sub-volume $dV = d_x \cdot d_y \cdot d_z$. Because the defect concentration is low, we can always choose large values of $d_x$, $d_y$, and $d_z$ (comparable to the defect scattering distance shown in Fig. \ref{flux_distribution}). Hence, the material outside the $dV = d_x \cdot d_y \cdot d_z$ volume can be viewed as far away from the defect and therefore its thermal conductivity remains to be the value of the (defect-free) matrix material, $\kappa_m$. Despite the non-uniform thermal properties on a fine scale, the smaller scattering volume $dV$ exhibits an apparent overall thermal conductivity $\kappa_d$. Since $\kappa_d$ is essentially the coarse-grained, average thermal conductivity of the volume $d_x \cdot d_y \cdot d_z$, it therefore depends strongly on $d_x$, $d_y$, and $d_z$; however, $\kappa_d$ can be viewed as a constant for a given defect size and type once $d_x$, $d_y$, and $d_z$ are given.

Using the model shown in Fig. \ref{model}(a), we define two geometric parameters $\alpha = \left(d_y \cdot d_z\right) / (\delta_y \cdot \delta_z)$ and $\beta = d_x / \delta_x$. Considering that $\delta_x$, $\delta_y$, and $\delta_z$ are the overall material dimensions divided by numbers of defects in the three coordinate directions, and $d_x$, $d_y$, and $d_z$ are given constants physically representing the dimension around the defect where scattering is significant, parameters $\alpha$ and $\beta$ prescribe the relative defect densities (areal and lineal fractions) in the cross-sectional and axial dimensions respectively. In particular, $\alpha$ and $\beta$ can be termed defect ``scattering'' densities because $\alpha = 1$ and $\beta = 1$ do not represent 100\% defect volume fraction (or site fraction), but rather indicate that the ratio between defect scattering volume and total volume equals one. We can also define a defect volume scattering density $\rho = \alpha \cdot \beta$. Note that for a given defect size and type, the defect scattering density $\rho$ is proportional to the defect volume fraction $\xi$, $\xi = f \cdot \rho$, with the scaling factor $f$ defined as
\begin{equation}
f = \frac{\Omega}{\Omega_s} = \frac{\ell_x \cdot \ell_y \cdot \ell_z}{\left(\ell_x + \ell_s\right)\cdot\left(\ell_y + \ell_s\right)\cdot\left(\ell_z + \ell_s\right)}
\label{conversion factor}
\end{equation}
where $\Omega$ is physical volume of defect, $\Omega_s$ represents a scattering volume, $\ell_x$, $\ell_y$, and $\ell_z$ are physical dimensions of the defect in the $x-$, $y-$, and $z-$ directions, and $\ell_s$ is a mean scattering distance for all phonons contributing to the heat flux. Eq. (\ref{conversion factor}) indicates that $f$ depends on defect size. As will be clear below, this is an important concept illustrating why thermal conductivity is sensitive to defect sizes.
 
Our distinction of two defect densities $\alpha$ and $\beta$ may not appear critical for uniform defect distributions. These two separate measures of defect density, however, are useful when the defect distribution is not uniform. For instance, consider the case of heat conduction through a long nanowire that is almost defect-free along its axial direction ($\beta \approx 0$) but is severely damaged or even completely fractured on one cross-section ($\alpha \approx 1$). Such a nanowire is expected to have a low thermal conductivity. On the other hand, if the same amount of defects is uniformly distributed along the axis of a long nanowire, defect densities become low in both cross-sectional and axial dimensions ($\alpha \approx 0$, $\beta \approx 0$). Such a nanowire is expected to have a thermal conductivity close to that of the matrix, $\kappa_m$. Clearly, the two parameters $\alpha$ and $\beta$ can be used to distinguish the two cases.

\subsection{Derivation of Thermal Conductivity as a Function of Defect Density, Size, and Distribution}

Analytical expression can be derived for thermal conductivity of a material as a function of defect density, size, and distribution using the coarse-grained conductivity concept just introduced. The free parameters of this model are the matrix and defect volume conductivities, $\kappa_m$ and $\kappa_d$, and the chosen defect volume size $d_x$, $d_y$, and $d_z$. First, the material volume $\delta_x \cdot \delta_y \cdot \delta_z$ is divided along the $x$ direction into a central rectangular prism with a cross-section area $d_y \cdot d_z$ that contains the defect and the surrounding material that is far away from the defect, as shown in the left frame of Fig. \ref{model}(b). The apparent overall thermal conductivity of the central prism can be assumed to be $\kappa_{d_yd_z}$, whereas that of the remaining material is $\kappa_m$. Because the heat conducts through the two parts of the material in parallel (on average), the thermal conductivity of the entire material (matrix + defect) can be calculated as an area weighted average \cite{ZJA2010a,ZJA2010b}:
\begin{equation}
\kappa_{m+d} = \frac{d_y \cdot d_z}{\delta_y \cdot \delta_z} \cdot \kappa_{d_yd_z} + \frac{\delta_y \cdot \delta_z - d_y \cdot d_z}{\delta_y \delta_z} \cdot \kappa_m = \alpha \cdot \kappa_{d_yd_z} + \left(1 - \alpha\right) \cdot \kappa_m
\label{kappa1}
\end{equation}
In Eq. (\ref{kappa1}), $\kappa_{d_yd_z}$ can be expanded by observing that the central prism is composed of a central section with a length $d_x$ and a conductivity $\kappa_d$ and two end sections with a total length $\delta_x - d_x$ and a conductivity $\kappa_m$ as shown in the right frame of Fig. \ref{model}(b). Because heat conducts through the three sections in serial, the overall thermal resistivity is calculated as a length weighted average \cite{ZJA2010a,ZJA2010b}:
\begin{equation}
\frac{1}{\kappa _{d_yd_z}}= \frac{d_x}{\delta_x} \cdot \frac{1}{\kappa_d} + \frac{\delta_x - d_x}{\delta_x} \cdot \frac{1}{\kappa_m} = \frac{\beta}{\kappa_d} + \frac{1-\beta}{\kappa_m}
\label{kappa dydz}
\end{equation}
Substituting Eq. (\ref{kappa dydz}) into Eq. (\ref{kappa1}), we have:
\begin{equation}
\kappa_{m+d} = \kappa_m - \kappa_m \cdot \left[1 - \frac{1}{\beta \cdot \left(\kappa_m - \kappa_d\right)/\kappa_d + 1}\right] \cdot \alpha = \kappa_m - \kappa_m \cdot \frac{\left(\kappa_m-\kappa_d\right)/\kappa_d}{\beta \cdot \left(\kappa_m - \kappa_d\right)/\kappa_d + 1} \cdot \rho 
\label{kappa2}
\end{equation}
Eq. (\ref{kappa2}) correctly predicts a small thermal conductivity of $\kappa_{m+d} \approx 0$ at $\alpha = 1$, $\beta \approx 0$, and $\kappa_d = 0$ (i.e., the cross-section is completely fractured), and a large thermal conductivity of $\kappa_{m+d} \approx \kappa_m$ at $\alpha \approx 0$, $\beta \approx 0$ and $\kappa_d \neq 0$.

Eq. (\ref{kappa2}) can be simplified for common scenario where defect distribution is uniform and defect concentration is low. In such cases, $\kappa_d \neq 0$ (note that the $\kappa_d = 0$ assumption used above implies a complete fracture of a cross-section normal to the heat flux direction). When large $d_x$, $d_y$ and $d_z$ values are used to contain the defects, we can always approach the limits $\kappa_d \approx \kappa_m$ and $\beta \cdot \left(\kappa_m - \kappa_d\right)/\kappa_d \approx 0$. Using the relation $1/\left(1 + x\right) \approx 1 - x$ for small $x$, Eq. (\ref{kappa2}) can be approximated as
\begin{equation}
\kappa_{m+d} \approx \kappa_m - \left(\kappa_m - \kappa_d\right) \cdot \alpha \cdot \beta = \kappa_m \left[1- \left(\kappa_m-\kappa_d\right)/\kappa_m \cdot \rho\right] = \kappa_m \left(1-\eta \cdot \xi\right)
\label{kappa2 simple}
\end{equation}
where $\eta = f^{-1} \cdot \left(\kappa_m-\kappa_d\right)/\kappa_m$ is a constant for a given type of defect with a given size and given pattern (e.g., array). Eq. (\ref{kappa2 simple}) indicates that thermal conductivity is a linear function of defect volume scattering density $\rho$. This verifies that for independent uniform defect distributions, distinction between areal and axial defect densities is not necessary. Furthermore, Eq. (\ref{kappa2 simple}) also indicates that thermal conductivity is a linear function of defect volume fraction $\xi$, for a given defect size, by way of the volume ratio $f$, Eq. (\ref{conversion factor}). Eq. (\ref{kappa2 simple}) can describe the scaling for the defect distribution shown in Fig. \ref{defect arrays}(a) if defect density is changed by the defect spacing $\delta$. It can also be used to describe the scaling for the defect distributions shown in Figs. \ref{defect arrays}(b) and Fig. \ref{defect arrays}(c) if the defect arrays remain unchanged (i.e., $\iota_x$, $\iota_y$, $\iota_z$ are kept constant) and the defect density is only changed by the spacing $\delta$ between arrays. In the latter case, we are studying the scaling of arrays rather than the scaling of individual defect.

Eq. (\ref{kappa2 simple}) can also be derived based upon Fig. \ref{model}(c) as is described in Appendix \ref{2nd derivation}. 

\section{Molecular Dynamics Verification}

In this section, we verify the model by performing large scale MD simulations (400 processors or more) using the bulk configuration (periodic boundary condition in the $y-$ direction). Two series of MD simulations are conducted to explore effects of the areal and axial defect densities respectively. For areal effects, we use the configuration shown in Fig. \ref{defect arrays}(b) where defect spacings in the $x-$ and $z-$ directions are small but are kept fixed whereas spacing between the $x-z$ defect arrays in the $y-$ direction is large and varied to change defect areal density. For the axial effects, we use the configuration shown in Fig. \ref{defect arrays}(c) where defect spacings in the $y-$ and $z-$ directions are small but are kept fixed whereas spacing between the $y-z$ defect arrays in the $x-$ direction is large and varied to change defect axial density. Because only the spacing in the direction with a sparse defect distribution is varied, this study satisfies our model assumption and can be used to test the model. The relatively small defect spacings in the other two directions enable small systems to be used to significantly reduce the computational expenses. In particular, all of our simulations used a small fixed dimension of $n_3 = 6$ (W $\approx 19 \AA$) in the $z-$ direction.

\subsection{Effects of Areal Defect Density}

To explore the areal defect density effect, our first series of simulations employ a fixed $x-$ dimension (aligned with the flow of heat) of $n_1 = 136$ ($2L \approx 707 \AA$), and various $y-$ dimensions between $n_2 = 30 - 100$ ($t \approx 166 - 553 \AA$). As a reference of the dimensions, the smallest and the largest systems contain respectively 195,840 and 652,800 atoms. A heat flux of $0.00035~eV/\AA^2 \cdot ps$ is used to introduce the temperature gradient. The choice of heat flux used in the work is to ensure that temperature difference between heat source and heat sink does not exceed 10 K but is significant enough to enable converged calculations. Four equal-spaced voids are introduced between heat source and sink (the system contains a pair of heat sources and sinks and therefore eight voids). Notice that the void spacing in the $x-$ direction does not affect the analysis of defect density in the $y-$ direction, nor does the particular choice of four voids affect the generality of the results. As has been well established in the past \cite{ZAJGS2009,ZJA2010b}, the use of different $y-$ dimensions above a certain threshold does not affect the thermal conductivity of a defect-free matrix with periodic boundary conditions in the lateral directions. It does affect the distance between periodic defects in the $y-$ direction (i.e., the spacing between the $x-z$ defect arrays) and hence the areal defect density. Clearly, the defect spacings $\delta_x$, $\delta_y$, and $\delta_z$ relate to system dimensions as $\delta_x = L/4$, $\delta_y = t$, and $\delta_z = W$. Since $L$ and $W$ are not varied (i.e., we are not exploring the scaling in the $x-$ and $z-$ directions), we can simply set the dimensions of the (four) scattering volumes in these directions to their maxima $d_x = L/4$ and $d_z = W$. A large $d_y$ value is desired to contain the defect scattering region in the $y-$ direction, but the geometry requires that $d_y \leq \delta_y$. We can maximize the constant $d_y$ by setting $d_y = \delta_{y,min} = t_{min}$, where $t_{min}$ is the minimum thickness of all the computational systems used in the series. This ensures that the geometry constraint is satisfied for all the samples. Under these conditions, the defect densities $\alpha = t_{min}/t$, $\beta = 1$, and $\rho = t_{min}/t$. Note that eight voids correspond to the removal of 96 atoms. For the smallest system, $\rho = 1$, and the defect volume fraction (site fraction) $\xi = 96/195840 \approx 4.9\times10^{-4} = 4.9\times10^{-2} \%$. This means that the defect volume fraction conversion factor for the series of samples is $f = \xi/\rho = 4.9\times10^{-2}$\%.

\begin{figure}
\includegraphics[width=6in]{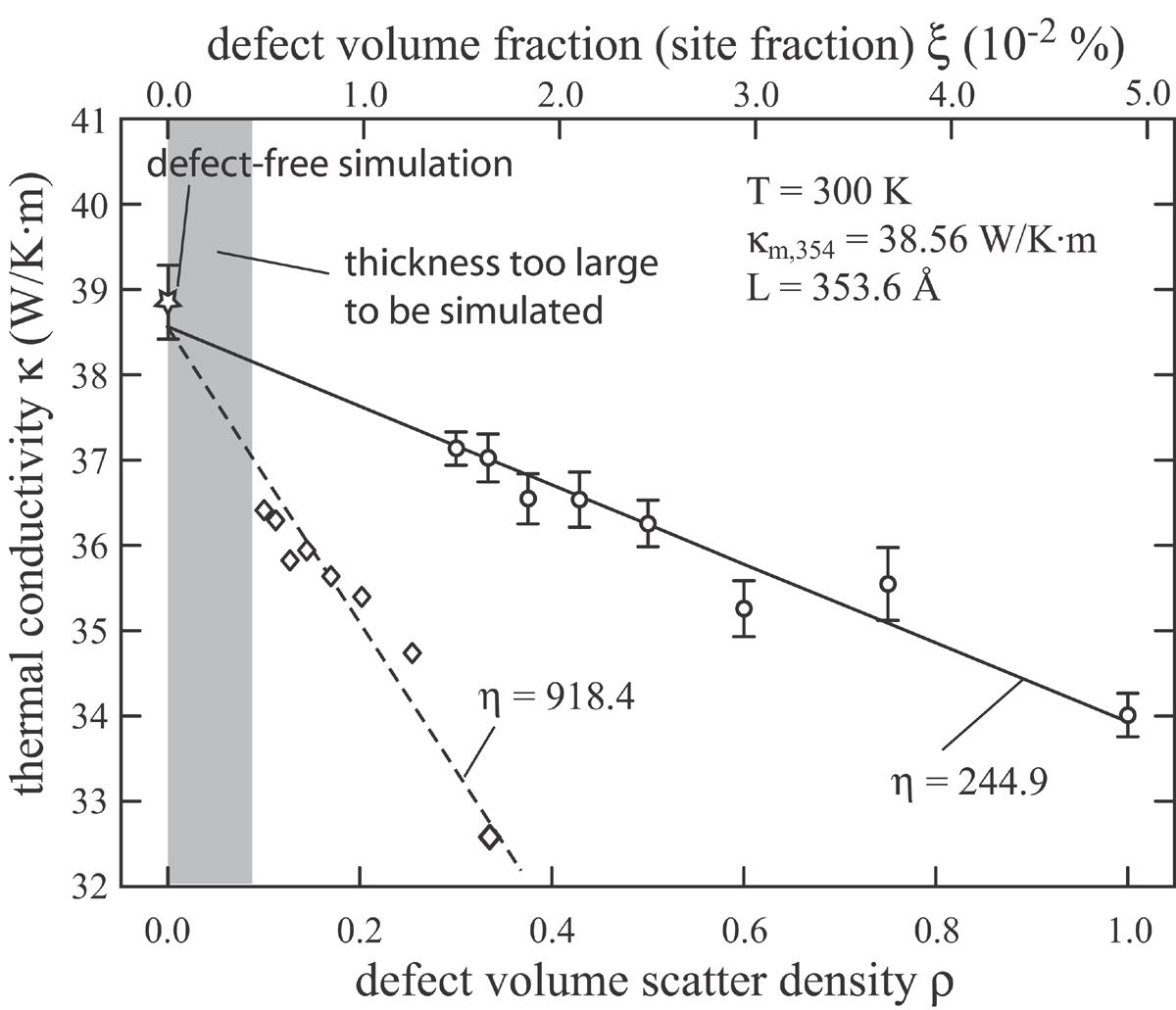}
\caption{Thermal conductivity as a function of the defect density $\rho$ or $\xi$. Unfilled circles and solid line correspond to planar void arrays spaced $88\times19$ \AA$^2$ in the $x-z$ planes, and unfilled diamonds and dashed line correspond to planar void arrays spaced $110\times19$ \AA$^2$ in the $y-z$ planes. Spacing between arrays is varied to change defect densities. Note that the unfilled diamonds are not the direct result of MD simulations (see Section \ref{sec:ax_dens}) and hence are not associated with error bars.
\label{kappa vs. alpha}}
\end{figure}

According to Eq. (\ref{kappa2 simple}) or (\ref{kappa2 simple1}), thermal conductivity is a linear function of $\rho$. This relationship can be verified using data from direct simulations which are shown in Fig. \ref{kappa vs. alpha} as a function of defect scattering density $\rho$ or defect volume fraction $\xi$ using unfilled circles. It should be noted that the shaded area corresponds to the low defect density regime that cannot be easily calculated via MD using current computers. The thermal conductivity at $\rho = 0$, however, can be estimated using a smaller, defect-free system with periodic boundary conditions. The value thus obtained is shown in Fig. \ref{kappa vs. alpha} with an unfilled star to distinguish it from other data points. The solid line in Fig. \ref{kappa vs. alpha} is a linear function fitted to the simulated data. Note that the dashed line and unfilled diamonds show the effects of the axial defect density obtained in another series of MD simulations, to be discussed in the following section. At this point, the most important result in Fig. \ref{kappa vs. alpha} is that the solid trend line fitted to the defect data \emph{predicts} closely the defect-free value. This is a strong verification of Eq. (\ref{kappa2 simple}) or (\ref{kappa2 simple1}). Also, since data at $\xi = 0$ and large $\xi$ can both be directly calculated using MD, Eq. (\ref{kappa2 simple}) or (\ref{kappa2 simple1}) can be used in an \emph{interpolation} to reliably predict thermal conductivities in the low defect density regime that is not directly accessible by MD simulations. This is unlike the extrapolation typically used to infer the bulk limit in the presence of size-effects \cite{ZAJGS2009}.

Numerically, our calculations indicated that the defect-free thermal conductivity at a sample length of $n_1 = 136$ ($2L \approx 707 \AA$) is $\kappa_{m,354} = 38.56~W/K \cdot m$ (see Fig. \ref{kappa vs. alpha}). Note that thermal conductivity depends on both defects and sample length \cite{M1992,PB1994,OS1999,ZJA2010a,ZJA2010b,ZAJGS2009}, and as a result, we use $\kappa_{m,354}$ instead of $\kappa_{m}$ to indicate that the matrix thermal conductivity determined here pertains specifically to a sample length of $L \approx 354~\AA$. The void configurations discussed here are essentially planar void arrays spaced $88\times19$ \AA$^2$ in the $x-z$ planes with spacing between arrays being varied between 166 and 553 \AA. For such particular defect configurations, we obtained a (non-dimensional) scaling coefficient of $\eta = 244.9$ for the effect of defect volume fraction on thermal conductivity.

\subsection{Effects of Axial Defect Density} \label{sec:ax_dens}

To explore the axial defect density effect, our second simulation series employs a fixed $y-$ dimension (perpendicular to the flow of heat) of $n_2 = 20$ ($t \approx 110 \AA$), and various $x-$ dimension between $n_1 = 152$ and $500$ ($2L \approx 790 - 2600 \AA$). As a reference of the dimensions, the smallest and the largest systems contain respectively 145,920 and 480,000 atoms. A heat flux of $0.0002 eV/\AA^2 \cdot ps$ is used to introduce the temperature gradient. One void is created in the middle between the heat source and sink (two voids in the system). Unlike the areal case where the change of thermal conductivity due to the change of system thickness $t$ in the $y-$ dimension comes only from the change of the areal defect density $\alpha$ (scales with $1/t$), the change of thermal conductivity due to the change of system length $L$ in the $x-$ dimension comes from both: (a) the change of the axial defect density $\beta$ (scales with $1/L$) and, (b) the size-dependence of the interfacial scattering \cite{M1992,PB1994,OS1999,ZJA2010a,ZJA2010b,ZAJGS2009}. Our scaling model can be used to derive an analytical expression of thermal conductivity as a function of both defect density and sample length, as described by Eq. (\ref{1 over defect kappa1}) in Appendix \ref{length conversion}. Based upon the relation $\rho \propto 1/L$, Eq. (\ref{1 over defect kappa1}) indicates that thermal resistivity of defective material with a finite length $L$ is a linear function of $1/L$. The validity of the analytical model on the axial defect density can therefore be verified by checking this linear relationship. Furthermore, since the length scaling coefficient $p$ (refer to Eq. \ref{1 over defect kappa1}) can be determined from a series of defect-free samples with different lengths \cite{ZAJGS2009}, the effect of defects can be isolated. 

\begin{figure}
\includegraphics[width=6in]{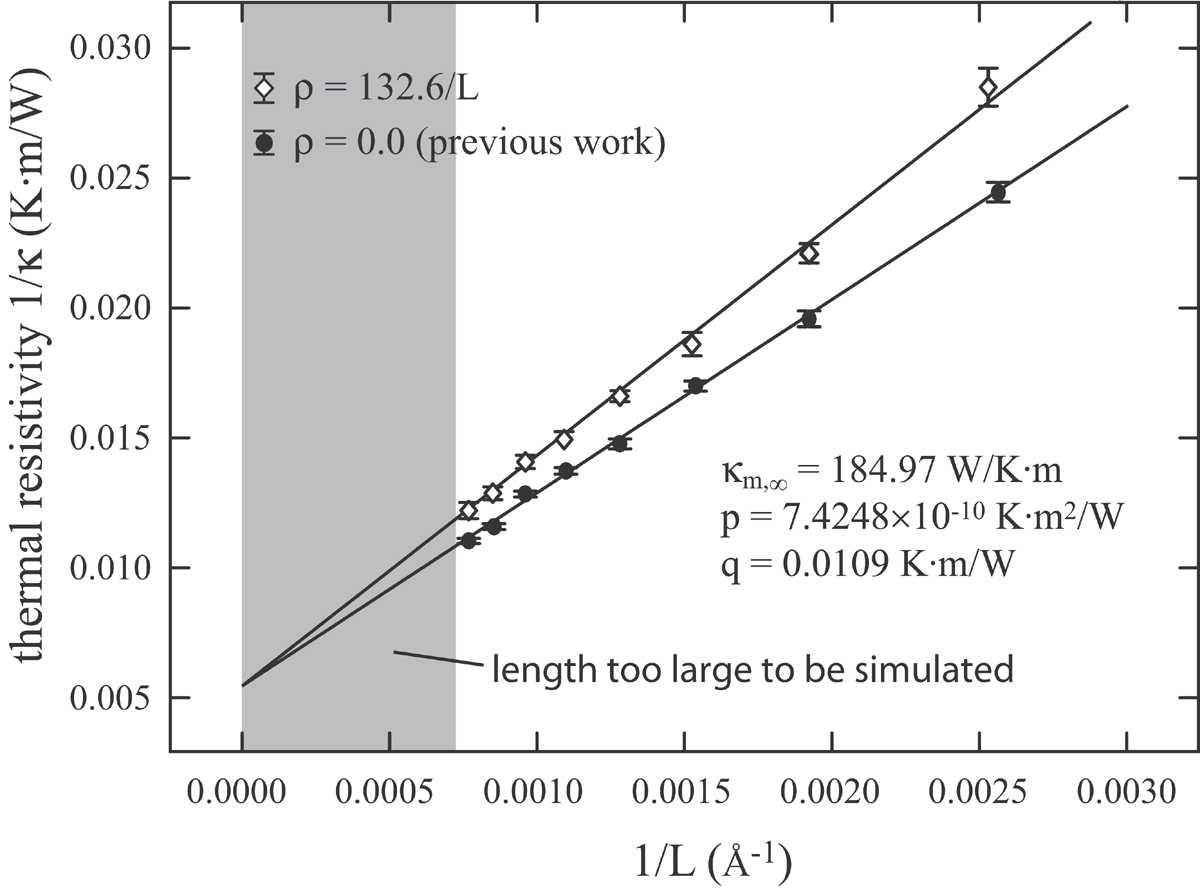}
\caption{Thermal resistivity as a function of inverse of sample length $1/L$ with and without the defect in the middle of sample.
\label{kappa vs. beta}}
\end{figure}

Here, the simulated thermal resistivity results are shown in Fig. \ref{kappa vs. beta} as a function of $1/L$ using the unfilled diamonds. For comparison, similar data for defect-free samples obtained previously \cite{ZAJGS2009} are included as the filled circles. Note that the lines are calculated from a function fitted to Eq. (\ref{1 over defect kappa1}) as will be described in the following. It can be seen that the linear relations are very well satisfied. Most importantly, the thermal conductivity data for the defective samples is seen to be lower than that of the defect-free samples, and both defective and defect-free samples approach the \emph{same} $\kappa_{m,\infty}$ limit at $L \rightarrow \infty$ (also $\rho \rightarrow 0$). As described in Appendix \ref{length scaling}, a direct test of how good the linear relationship extends to the infinite sample length is not possible using merely defect-free samples because it is increasingly difficult to obtain accurate thermal conductivities from MD simulations when sample length is increased. However, the convergence of defective and defect-free samples to the same thermal conductivity at the infinite sample length limit is a strong verification of the linear scaling law, Eq. (\ref{1 over defect kappa1}). Eq. (\ref{1 over defect kappa1}) therefore can accurately predict thermal conductivities in the defect density and sample dimension space that cannot be directly assessable by MD. 

Now, we compare the results of this section to those of the previous section which examined the influence of areal density. Here, we set $d_y = \delta_y = t$, $d_z = \delta_z = W$. In the previous section, a defect volume scattering density $\rho = 1$ corresponds to eight voids in a $136\times30\times6$ cells$^2$ system. This density is equivalent to two voids in a $51\times20\times6$ cells$^2$ system. To match the previous defect volume scattering density definition, we hence choose $d_x = 51/2$ cells $= 132.6$ \AA~($< \delta_{x,min} = L_{min}$, where $L_{min}$ corresponds to the minimum $x-$ dimension per defect used in the simulation series). At these values, the defect densities $\alpha = 1$, $\beta = 132.6/L$, and $\rho = 132.6/L$. Using these definitions, we fit Eq. (\ref{1 over defect kappa1}) to our data. This reproduced the bulk limit of the thermal conductivity value of $\kappa_{m,\infty} = 184.97~W/K \cdot m$ obtained previously \cite{ZAJGS2009}, and resulted in the determination of a length scaling coefficient of $p = 7.4248 \times 10^{-10}~K \cdot m^2/W$, and a defect volume scattering density scaling coefficient of $q = 0.0109~K \cdot m/W$, see Eq. (\ref{1 over defect kappa1}). The lines shown in Fig. \ref{kappa vs. beta} are calculated using these parameters. Note that in this series of simulations, the defects are essentially planar arrays of voids spaced $110\times19$ \AA$^2$ in the $y-z$ planes with array spacing being varied between 395 and 1300 \AA.

Eq. (\ref{1 over defect kappa1}) also allows us to cast the thermal conductivity obtained at one sample length $L_1$ to another $L_2$ at a constant defect density:
\begin{equation} 
\frac{1}{\kappa_{m+d}\left(\rho\right)|_{L_2}} = \frac{1}{\kappa_{m+d}\left(\rho\right)|_{L_1}} +p \cdot \left(\frac{1}{L_1} - \frac{1}{L_2}\right)
\label{recast kappa vs. beta}
\end{equation} 
Using Eq. (\ref{recast kappa vs. beta}), thermal conductivity vs. defect density data at a fixed sample length of $n_1 = 136$ ($2L \approx 707$ \AA) is obtained. Fitting Eq. (\ref{kappa2 simple}) or (\ref{kappa2 simple1}) to such data resulted in $\kappa_{m,345} = 38.56~W/K \cdot m$ and a scattering coefficient of $\eta = 918.4$ for the effect of defect volume fraction on thermal conductivity. Pertaining to the same sample length as the unfilled circles in Fig. \ref{kappa vs. alpha}, the converted data and the corresponding fitted function are shown as unfilled diamonds and dash line in Fig. \ref{kappa vs. alpha}. It can be seen that like the areal defect density effect, reducing the axial defect density also causes the thermal conductivity to approach the value of the defect-free sample. With consideration of Appendix \ref{length scaling} which discusses the difficulties in directly testing the thermal conductivity at very large sample length, the convergence to the same defect-free sample point from two defect relations shown in Fig. \ref{kappa vs. alpha} is again a strong verification of the analytical model.

For non-zero defect densities, thermal conductivities obtained from the areal and axial series are different, with different coefficients $\eta$. This means that when defects are not sparse, their scattering regions can overlap, resulting in different effects on thermal conductivities depending on how defects are distributed. For example, the configurations shown in Figs. \ref{defect arrays}(b) and \ref{defect arrays}(c) have different scaling coefficients.

\section{Effect of Defect Spatial Distribution}

\begin{figure}
\includegraphics[width=6in]{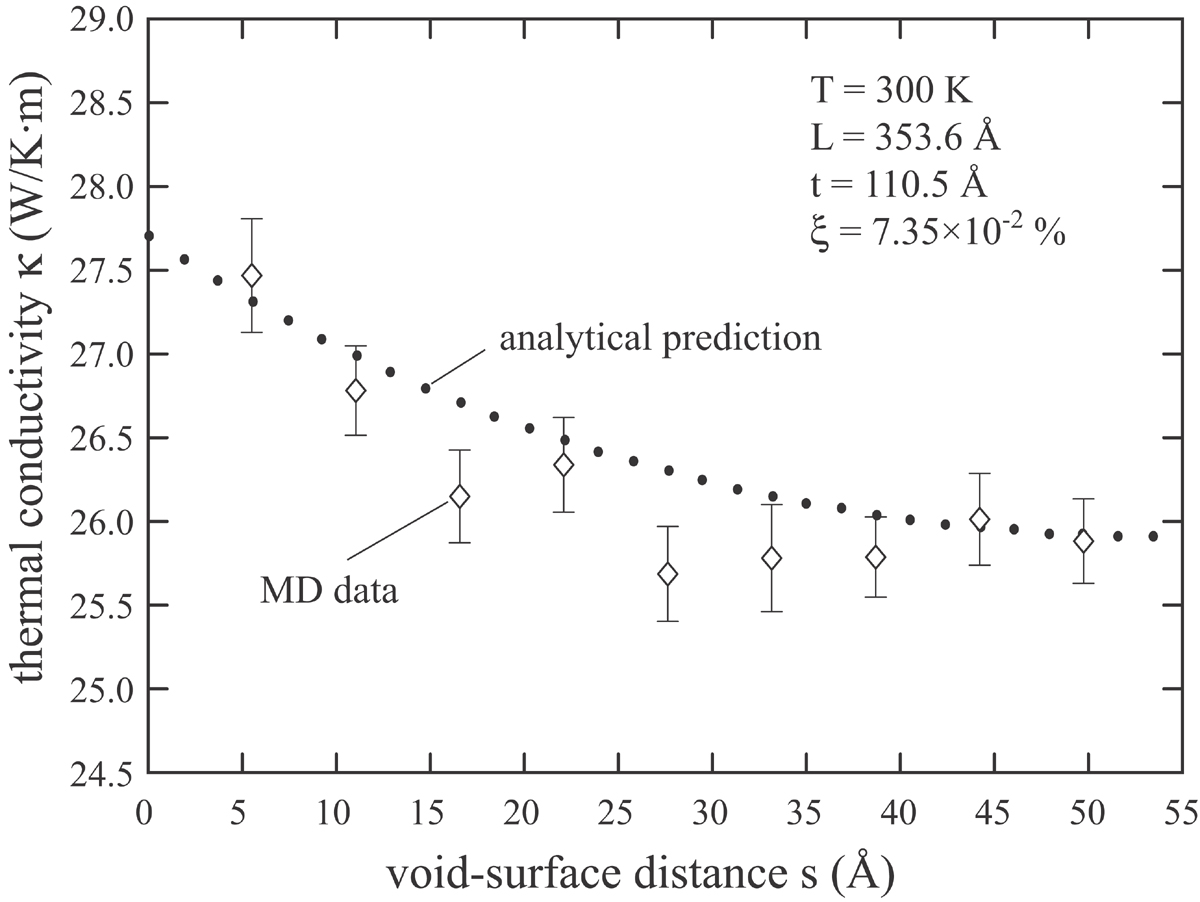}
\caption{Thermal conductivity of a thin film as a function of void-surface distance predicted by MD simulations and an analytical model.
\label{offset}}
\end{figure}

A series of MD simulations are performed to study the effect of defect location with respect to a surface using a film configuration with the free boundary condition in the $y-$ direction. The system has a fixed dimension of $n_1 = 136$ ($2L \approx 707$ \AA), $n_2 = 20$ ($t \approx 110 \AA$), and $n_3 = 6$ (W $\approx 19 \AA$) for a total of 130,560 atoms. Four voids are created between the heat source and sink, resulting in a void volume fraction of $\xi = 7.35\times10^{-2}~\%$. These defects are uniformly distributed in the $x-$ direction (aligned with the heat flow), but the position of the row of defects is varied along the $y-$ direction so that it has different distance $s$ from the free surface, Fig. \ref{MD geometry}. The calculated thermal conductivities as a function of $s$ are shown with unfilled diamonds in Fig. \ref{offset}. Fig. \ref{offset} clearly indicates that thermal conductivity depends on the defect-surface distance. In particular, thermal conductivity increases as defects move towards the surface.

Analysis can be used to understand the results shown in Fig. \ref{offset}. Eq. (\ref{kappa2 simple}) or (\ref{kappa2 simple1}) indicates that if the total scattering strength of one type of defects is $\eta_1$ (aggregating the dependence on the defect volume fraction $\xi$), then the thermal conductivity $\kappa_1$ is reduced from the matrix value $\kappa_0$ through $\kappa_1 = \kappa_0 \cdot \left(1-\eta_1\right)$. Similarly, if there is a second type of defects with the scattering strength $\eta_2$, then the thermal conductivity $\kappa_2$ of the material can be viewed as reduced from the new matrix value $\kappa_1$ through $\kappa_2 = \kappa_1 \cdot \left(1-\eta_2\right) = \kappa_0 \cdot \left(1-\eta_1\right) \cdot \left(1-\eta_2\right)$. In general, we can assume a multiplication rule of $\kappa_N = \kappa_0 \cdot \Pi_{i=1}^N \left(1-\eta_i\right)$ to account for $N$ types of defects. To relate to the MD results, a two dimensional illustration of the system is shown in Fig. \ref{defect population model}, where voids are assumed to lie on the center line of the shaded region a distance $s$ below the surface, and the total sample thickness is $t$. Because the interaction between surface and defects needs to be addressed, the thermal conductivity must be considered locally as a function of position y. 

It is obvious from Fig. \ref{flux_distribution} that the local scattering strength $\eta_{l,s}\left(d\right)$ of a surface is a decreasing function of distance $d$ from the surface. We postulate that such a behavior can be well captured by the function:
\begin{figure}
\includegraphics[width=6in]{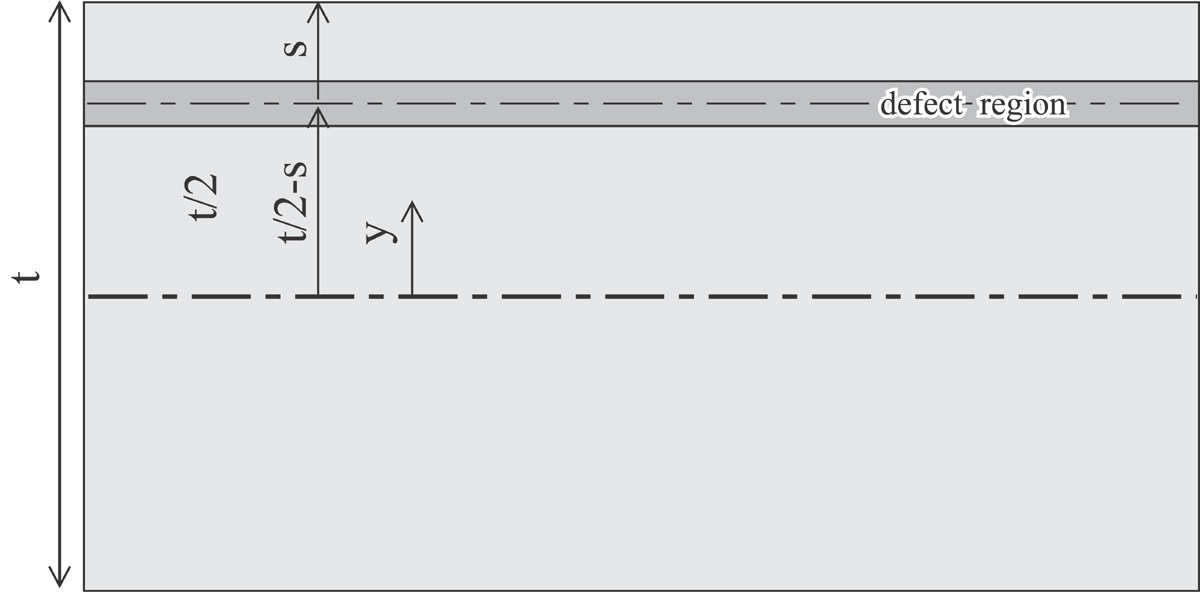}
\caption{Illustration of defect population model.
\label{defect population model}}
\end{figure}
\begin{equation} 
\eta_{l,s}\left(d\right) = \delta_{\kappa,s} \cdot \exp\left(-\mu \cdot d\right)
\label{eta_l,s}
\end{equation} 
where $\delta_{\kappa,s}$ and $\mu$ are constants. The local thermal conductivity $\kappa_{l,s}\left(y\right)$ due to the presence of the surfaces (but no defects) is therefore: 
\begin{equation} 
\kappa_{l,s}\left(y\right) = \kappa_m \cdot \left[1-\eta_{l,s}\left(t/2-y\right)\right] \cdot \left[1-\eta_{l,s}\left(t/2+y\right)\right]
\label{kappa_l,s}
\end{equation} 
where $t/2 \pm y$ are the distances of a local site $y$ from the two surfaces. Similarly, we can assume that the local scattering strength $\eta_{l,d}\left(d\right)$ of defects is a decreasing function of distance $d$ from the defect plane based upon Fig. \ref{flux_distribution}. Again, we postulate that the behavior can be well captured by the function:
\begin{equation} 
\eta_{l,d}\left(d\right) = \delta_{\kappa,d} \cdot \exp\left(-\nu \cdot d\right)
\label{eta_l,d}
\end{equation} 
where $\delta_{\kappa,d}$ and $\nu$ are constants. The local thermal conductivity $\kappa_{l,d}\left(y\right)$ due to the presence of the defects (but no surface) is then
\begin{equation}
\kappa_{l,d}\left(y\right) = \kappa_m \cdot \left[1-\eta_{l,d}\left(\left|t/2-s-y\right|\right)\right]
\label{kappa_l,d}
\end{equation}
where $\left|t/2-s-y\right|$ measures the distance of a local site $y$ from the defect plane. Note that Eq. (\ref{kappa_l,d}) prescribes the ``apparent'' thermal conductivity in a thin slice parallel to the defect plane at a location of $y$. For such a slice, Eq. (\ref{kappa_l,d}) correctly specifies a drop of the thermal conductivity by a fraction of $\delta_{\kappa,d}$ at the defect plane $y = t/2-s$, and the recover of the bulk value $\kappa_m$ when $y$ is far away from the defect plane with the parameter $\nu$ essentially capturing the scattering distance.

In regions where both surface and defects have significant scattering, the total local thermal conductivity $\kappa_l\left(y\right)$ can be written as: 
\begin{equation} 
\kappa_{l}\left(y\right) = \kappa_m \cdot \left[1-\eta_{l,s}\left(t/2-y\right)\right] \cdot \left[1-\eta_{l,s}\left(t/2+y\right)\right] \cdot \left[1-\eta_{l,d}\left(\left|t/2-s-y\right|\right)\right]
\label{kappa_l}
\end{equation} 
The apparent thermal conductivity of the system can then be found by averaging Eq. (\ref{kappa_l}) as
\begin{equation}
\kappa = \frac{1}{t} \int_{y=-\frac{t}{2}}^{y=\frac{t}{2}} \kappa_{l}\left(y\right) dy
\label{kappa y population}
\end{equation}

To apply Eq. (\ref{kappa y population}), parameters $\kappa_m$, $\mu$, $\nu$, $\delta_{\kappa,s}$, and $\delta_{\kappa,d}$ are needed. For the simulated sample length, the matrix conductivity is $\kappa_m = \kappa_{m,354} = 38.56~W/K \cdot m$. The other parameters can be estimated from independent simulations. First, we express the average thermal conductivity of a system containing the surfaces but not the defects as
\begin{equation}
\kappa_{s} = \frac{2}{t} \int_{y=0}^{y=t/2} \kappa_{l,s}\left(y\right)dy
\label{kappa s}
\end{equation}
Here we only integrate half of the sample thickness due to the symmetry of the problem. A series of MD simulations are carried out to obtain $\kappa_s$ vs. $t$ data for defect-free thin film samples. In particular, we use a fixed $x-$ dimension of $n_1 = 136$ ($2L \approx 707 \AA$) and various $y-$ dimensions of $n_2 = 30, 50, 70, 96$ with free $y-$ surfaces ($t \approx 166 - 530 \AA$). In addition, we perform an additional simulation with the periodic condition in the $y-$ direction at $n_2 = 10$ ($t \rightarrow \infty$). By fitting Eq. (\ref{kappa s}) to the $\kappa_s$ vs. $t$ results obtained from MD simulations, we find $\delta_{\kappa,s} = 0.15135$ and $\mu = 0.00705$ \AA$^{-1}$.

The average thermal conductivity of a periodic (i.e., no surfaces) system containing defects on the center line (i.e., $s = t/2$) is expressed as
\begin{equation}
\kappa_{d} = \frac{2}{t} \int_{y=0}^{y=t/2} \kappa_{l,d}\left(y\right)dy
\label{kappa d}
\end{equation}
Note that the defect scattering densities used in Fig. \ref{kappa vs. alpha} (unfilled circles) satisfy $\beta = 1$ and $\rho = t_{min}/t$, where $t_{min} = 166$ \AA~($n_2 = 30$) is the minimum system thickness used in the series. Hence, the thermal conductivity vs. defect density ($\rho$) data shown in Fig. \ref{kappa vs. alpha} can be converted to the thermal conductivity vs. $1/t$ data. By fitting Eq. (\ref{kappa d}) to these conductivity vs. $1/t$ data, we obtain $\delta_{\kappa,d} = 0.29017$ and $\nu = 0.02759$ \AA$^{-1}$.

Based upon the parameters thus obtained, Eq. (\ref{kappa y population}) is used to calculate thermal conductivity as a function of $s$, and the results are included as the dots in Fig. \ref{offset}. A good agreement between the analytical prediction and the MD data is clearly shown, thereby verifying the analytical model. The mechanism of the defect spatial effects is now clear. Eq. (\ref{kappa_l}) indicates that when $s \rightarrow 0$, the defect affected region merges with one of the surface affected regions. Consequently, the total scattering affected region is reduced from the case where defect is in the middle of the sample ($s \rightarrow t/2$). It is this consolidation of the defect and surface scattering that causes an increase in thermal conductivity when defects approach the surface.

\section{Discussion}

Thermal conductivity as a function of void density, size, and distribution have not been experimentally measured. However, analytical expressions for thermal conductivity as a function of porosity (i.e., volume fraction $\xi$) have been experimentally derived for a variety of porous materials \cite{SC2004,BBS1997,R1996,BBMMR2005,R1975}. While different forms of analytical expressions are used to enable good fit of experimental data up to a high porosity of 0.8 \cite{R1975}, all expressions can be reduced accurately to a linear function $\kappa_{m+d} = \kappa_m \cdot \left(1 - \eta \cdot \xi\right)$ when porosity $\xi$ is small enough (below 0.05). This is in good agreement with Eq. (\ref{kappa2 simple}) or (\ref{kappa2 simple1}), which is also valid for void volume fraction far less than 0.05. However, there is a significant difference between the nanoscale void effects and the macro-scale porosity effects as the scaling parameter $\eta$ for different porous materials falls between 1.0 and 4.6 \cite{SC2004,BBS1997,R1996,BBMMR2005,R1975} whereas $\eta$ is 244.9 and 918.4 for the two void configurations explored in Fig. \ref{kappa vs. alpha}. This means that the thermal conductivity at a given void density cannot be interpolated linearly between a defect-free sample and porous materials, i.e. as void size changes. 

The discussion of the spatial effects of defects presented in the preceding section indicates that thermal conductivity increases when defects move closer because their scattering regions overlap resulting in a reduction of total scattering volume. This can account for the difference between voids and pores, which can be more clearly illustrated using Eq. (\ref{conversion factor}). When a large number of voids are closely packed to form a pore, the pore size $\ell_x$, $\ell_y$, and $\ell_z$ become very large ($> 2\mu m$ \cite{SC2004}) but the scattering dimension $\ell_s$ remains small (assumed to be comparable to that in the void case). This means that the defect volume-to-scattering density conversion factor $f$ approaches 1 for large pores. If the thermal conductivity $\kappa_d$ of a pore approaches 0, $\eta$ approaches 1 by definition. This explains why the parameter $\eta$ for a variety of porous materials falls in the order of 1. On the other hand, if a large pore is split into a large number of voids of sizes around 5 \AA~distributed uniformly in the material, a large number of independent scattering volumes will be created, resulting in a significant reduction of the conversion factor $f$, a significant increase in the parameter $\eta$, and hence a significant reduction of thermal conductivity. This accounts well with the previous MD results that vacancies cause a more significant reduction of thermal conductivity than voids given the same defect site fraction \cite{SGOS2011}. Interestingly, experiments also indicate that at a given porosity, reducing pore sizes causes an increase in $\eta$ \cite{SC2004}, which is consistent with an increase in total scattering volume. Our analytical model and MD data, therefore, are well corroborated by the experimental data on porous materials. 

At the lower limit of defect size, point defects have been studied extensively and the low temperature effect of Rayleigh scattering is well-known. However, there have been relatively few experimental studies of actual vacancies - most experimental data is for isotopic substitutions. Che et al. \cite{CCDG2000} calculated the effects of point vacancy concentrations in the range 0.01-0.16\% for carbon. Unfortunately, the power-law-like empirical model they fitted to the data does not have finite derivative at a zero defect concentration. Nevertheless, it is clear that this limit of phonon scattering from voids results in a scaling coefficient significantly larger than the macroscopic porosity case.

Molecular dynamics simulations have been recently used to study thermal transport of nanoporous crystalline \cite{FP2011} and amorphous \cite{CFP2011} silicon. The results clearly indicate that thermal conductivity depends on pore size and pore fraction. In particular, thermal conductivity was found to reduce with an increasing interfacial area concentration, which relates well to the effects of the scattering volume discussed above. Our studies, therefore, verify the previous results. In addition to the pore size and pore fraction effects, we further show that thermal conductivity is sensitive to defect population (for example, areal and axial populations have different effects). 

\section{Conclusions}

An integrated approach combining a physically-motivated analytical model, large scale MD simulations, and extensive experimental thermal conductivity data of porous materials is used to study defect effects on thermal conductivity. Corroborated results lead to an explicit functional expression of thermal conductivity on defect density, size, and spatial population. The following conclusions are of particular interests to both theoretical thermal transport studies and phonon engineering of materials:

\begin{itemize}
\item[(a)] Thermal conductivity depends strongly on total scattering volume of defects. This scattering volume differs from the physical volume of defects. It, however, can be linearly correlated with physical volume of defects when defect type and size are fixed. It is the defect configuration that minimizes this scattering volume but not the physical volume that will increase thermal conductivity. 
\item[(b)] When defects are close, their scattering regions overlap, resulting in reduced total scattering volume. As a result, thermal conductivity increases when voids move towards surfaces, or small voids collapse to form large pores.
\item[(c)] For uniform, sparse defect distribution with a given defect size, thermal conductivity is a linear function of defect volume fraction. However, thermal conductivity at a given void density cannot be interpolated between defect-free samples and macroscopically porous materials due to the difference in defect sizes. In general, the dependence of thermal conductivity on defects is not simply through volume fraction, but strongly depends on the size distribution of the voids.
\item[(d)] The analytical model enables thermal conductivities obtained from molecular dynamics simulations to be extrapolated/interpolated reliably to realistic defect density ranges, as well as the defect-free limit.
\end{itemize}

Finally, we point out that the success of our approach can be related to the explicit incorporation of scattering of short wavelength and short mean free path phonons in direct molecular simulations. These modes are scattered the most given the general behavior of Rayleigh scattering and therefore their contribution to the overall conductivity is the most sensitive to changes in defect density. On the other hand, the longer wavelength modes are less affected by point-like defects and their contribution to the thermal conductivity is given by the scaling analysis that estimates the long sample length limit.

\begin{acknowledgments}

Sandia National Laboratories is a multi-program laboratory managed and operated by Sandia Corporation, a wholly owned subsidiary of Lockheed Martin Corporation, for the U.S. Department of Energy$'$s National Nuclear Security Administration under contract DE-AC04-94AL85000. This work was performed under a Laboratory Directed Research and Development (LDRD) project. We are grateful for helpful discussions with Patrick Hopkins, John Duda, and Jonathan A. Zimmerman.

\end{acknowledgments}
 
\appendix{}

\section{Examination of the Length Scaling Law}\label{length scaling}

Michalski \cite{M1992} first discovered that when a heat flux passes through a material with a length $L$, the inverse of thermal conductivity of the material, $1/\kappa$, linearly increases with $1/L$ (i.e., $1/\kappa$ = $1/\kappa_{\infty}$ + $p/L$ where $\kappa_{\infty}$ is the thermal conductivity at $L \rightarrow \infty$ limit and $p$ is a length independent slope parameter). This scaling law was later also reported by other researchers \cite{PB1994,OS1999}. Since then, such a law has been used to calculate $\kappa_{\infty}$ by linearly extrapolating the thermal conductivity data obtained in ``direct method'' MD simulations at a series of short sample lengths \cite{ZAJGS2009,SPK2002,YCSS2004}. However, controversy has arisen in recent studies regarding the validity of the linear extrapolation method. In fact, our MD simulations on GaN thermal transport using periodic boundary condition in the heat flux direction \cite{ZAJGS2009} indicated that while the linear relation holds well at low temperatures of 300 K and 500 K, it deviates at 800 K. In particular at 800 K , the slope increases discontinuously when the sample length is increased above a threshold value, indicating two regimes and resulting in a larger extrapolated thermal conductivity when the sample length range is above this threshold. Similar phenomenon was later observed by other researchers \cite{SLTMA2010,YTT2010,HDG2011,TIM2010}. Based on the studies of Si using non-periodic boundary conditions in the thermal transport direction and motivated by the concept that diffusive thermal transport does not hold at length-scales below the phonon mean free path, Sellen et al \cite{SLTMA2010} suggested that the linear extrapolation procedure might not be valid unless the sample lengths used in MD simulations are significantly longer than those commonly used in literature. Recently, we derived an analytical expression relating thermal conductivity to dimensions in all three coordinate directions, and verified using large scale molecular dynamics simulations that this expression is satisfied when the sample dimensions are above certain threshold values \cite{ZJA2010a,ZJA2010b}.

While the literature studies cited above involve some of the largest MD simulations performed so far in the field, they have not given a fully consistent story. For example, Sellen et al's work \cite{SLTMA2010} suggested that the extrapolated bulk thermal conductivity $\kappa_{\infty}$ is underestimated if the MD systems are too short, and that realistic $\kappa_{\infty}$ can eventually be achieved if the MD samples are sufficiently long. However, our previous work \cite{ZAJGS2009} indicated that at least for GaN at 800 K, unrealistically large $\kappa_{\infty}$ can be obtained using MD sample lengths in a range shorter than that proposed by Sellen et al's work \cite{SLTMA2010} and reasonable $\kappa_{\infty}$ are obtained only when sample length is further reduced. While the underlying mechanism for this abnormal phenomenon was not clear, we emphasize that our results were obtained from very long averaging times (at least 40 ns and some are significantly longer) and hence the unrealistically large extrapolated thermal conductivity was unlikely to be caused by statistical errors. On the other hand, Yang et al \cite{YTT2010} indicated that for Si with sample lengths up to $\sim$ 200 nm, the linear relation holds well under the periodic boundary condition in the thermal transport direction, but changes discontinuously under a non-periodic condition. Clearly, the sample length range within which the linear relation holds depends on the boundary conditions. To help elucidate the literature observations on the linear scaling law, here we re-examine effects of both the boundary condition and sample length by performing direct method MD simulations using averaging times and a sample length significantly beyond those used previously. The same approach used previously \cite{ZAJGS2009} is followed to study the perfect GaN system at a 300 K temperature.

\begin{figure}
\includegraphics[width=6in]{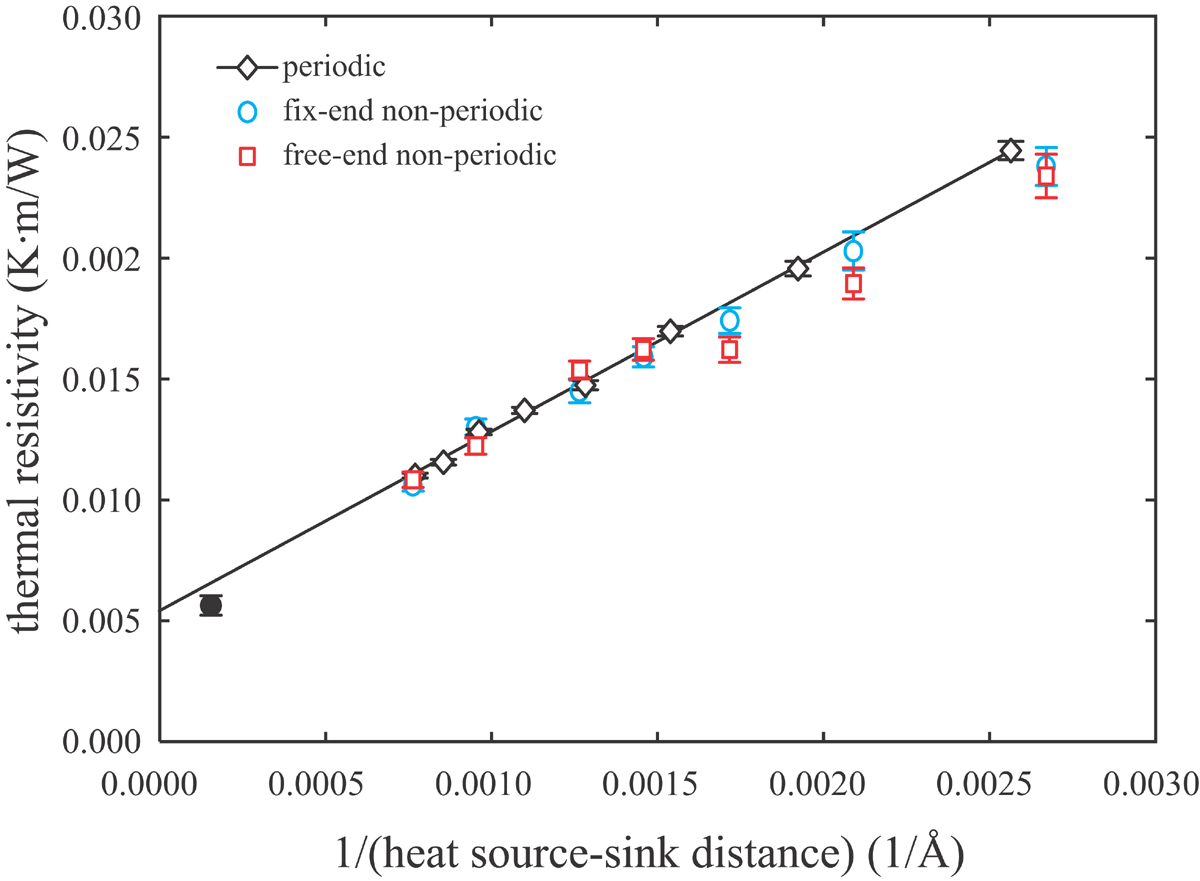}
\caption{Thermal resistivity of GaN crystals as a function of inverse of heat source-sink distance ($1/L$) at 300 K temperature. 
\label{Kvsl}}
\end{figure}

First, simulations are performed to calculate thermal conductivities of GaN crystals at different sample lengths. The crystal configurations are the same as those used previously \cite{ZAJGS2009} except that free boundary conditions are used in the thermal transport direction here whereas the previous work employed a periodic boundary condition. Two scenarios are considered, one with the two free-ends fixed, the other with the two free-ends free. To ensure highly-converged results, we use a very long averaging time of 110 ns. The results of thermal resistivity ($1/\kappa$) vs. inverse of thermal transport distance ($1/L$) obtained for the fixed and free ends are shown respectively as blue circles and red squares in Fig. \ref{Kvsl}. For comparison, the data obtained previously for the periodic boundary conditions \cite{ZAJGS2009} are included in Fig. \ref{Kvsl} using unfilled diamonds. Fig. \ref{Kvsl} clearly shows that the boundary condition has a significant effect on the linearity given the same sample length range. For the GaN system explored here, the free-end non-periodic condition has the most significant non-linearity, whereas the periodic condition produces the most linear relation.

One obvious question is how the phonon mean free paths of the collection of phonons affect the linear scaling law. The phonon mean free paths (and their associated relaxation times) are very difficult to calculate from MD simulations \cite{HC2008,STMA2010}, but the mean free path for GaN at 300 K has been experimentally measured to be around $0.1~\mu m$ \cite{DPWJP2006}. Hence, a GaN crystal with a sample length of $2L = 1.3~\mu m$ is used to perform an additional MD simulation under the periodic boundary condition. While the thermal transport distance $L$ equals half of the sample length under the periodic boundary condition and the phonon mean free path is uncertain under the simulated condition, it should be noted that our $L$ value is comfortably above the experimental phonon mean free path.

An increased sample length causes a significant increase in both the equilibration time for establishing temperature gradient and the averaging time for highly-converged results (at least scales with $L^2$). We hence perform a very long pre-MD run (64 ns) to equilibrate the temperature gradient, and use a very long (32 ns) simulation over which average temperature profiles are computed and used to evaluate the thermal conductivity. The result thus obtained is included in Fig. \ref{Kvsl} using the filled circle. It can be seen that the filled circle is very close to the solid line deduced from a linear expression of the MD data simulated under the periodic boundary condition in a small sample length range (i.e., the unfilled diamonds). Fig. \ref{Kvsl}, therefore, does not indicate that the length scaling law is violated for GaN crystals under periodic boundary conditions. This is a result that corroborates the findings in Yang et al \cite{YTT2010}. However, both our previous GaN work at 800 K \cite{ZAJGS2009} and the calculations here strongly indicate that converged thermal conductivity at a long sample length is extremely difficult to obtain. Note that the simulation time we use here is limited by the computing resources currently available to us and we only calculate one point at the very long sample length. While our MD time is significantly longer than that of most work reported in literature, we propose to significantly increase the averaging time in the future to re-calculate many points in a long sample length range when more powerful computing resources are available. 

\section{Second Derivation of Thermal Conductivity as a Function of Defect Density, Size, and Distribution} \label{2nd derivation}

First, the material volume $\delta_x \cdot \delta_y \cdot \delta_z$ is divided along the x direction into three slabs, the central slab has a thickness $d_x$ that contains the defect, and the remaining two end slabs have a total thickness of $\delta_x - d_x$, as shown in the left frame of Fig. \ref{model}(c). The apparent thermal conductivity of the central slab is assumed to be $\kappa_{d_x}$, and the thermal conductivity of the remaining two end slabs remains to be the matrix value $\kappa_m$. Because the heat conducts through the three slabs in serial, the overall thermal resistivity of the material can be represented as a length weighted average:
\begin{equation}
\frac{1}{\kappa_{m+d}} = \frac{d_x}{\delta_x} \cdot \frac{1}{\kappa_{d_x}} + \frac{\delta_x - d_x}{\delta_x} \cdot \frac{1}{\kappa_m} = \frac{\beta}{\kappa_{d_x}} + \frac{1-\beta}{\kappa_m}
\label{1 over kappa1}
\end{equation}
In Eq. (\ref{1 over kappa1}), $\kappa_{d_x}$ can be further expressed. This is illustrated in the right frame of Fig. \ref{model}(c), where the central slab $d_x \cdot \delta_y \cdot \delta_z$ is decomposed into a central volume $d_x \cdot d_y \cdot d_z$ and the remaining material. Because heat conducts through the central and remaining materials in parallel, the overall thermal conductivity of the central slab is an area weighted average:
\begin{equation}
\kappa_{dx} = \frac{d_y \cdot d_z}{\delta_y \cdot \delta_z} \cdot \kappa_d + \frac{\delta_y \cdot \delta_z - d_y \cdot d_z}{\delta_y \cdot \delta_z} \cdot \kappa_m = \alpha \cdot \kappa_d + \left(1-\alpha\right) \cdot \kappa_m
\label{kappa dx}
\end{equation}
Substituting Eq. (\ref{kappa dx}) into Eq. (\ref{1 over kappa1}), we have:
\begin{equation}
\frac{1}{\kappa_{m+d}} = \frac{1}{\kappa_m} - \frac{1}{\kappa_m} \cdot \left[1 - \frac{1}{\alpha \cdot \left(\kappa_d - \kappa_m\right)/\kappa_m + 1}\right] \cdot \beta = \frac{1}{\kappa_m} - \frac{1}{\kappa_m} \cdot \frac{\left(\kappa_d - \kappa_m\right)/\kappa_m}{\alpha \cdot \left(\kappa_d - \kappa_m\right)/\kappa_m + 1} \cdot \rho 
\label{1 over kappa2}
\end{equation}
Like Eq. (\ref{kappa2}), Eq. (\ref{1 over kappa2}) also correctly predicts a small thermal conductivity of $\kappa_{m+d} \approx 0$ at $\alpha = 1$, $\beta \approx 0$, and $\kappa_d = 0$, and a large thermal conductivity of $\kappa_{m+d} \approx \kappa_m$ at $\alpha \approx 0$, $\beta \approx 0$, and $\kappa_d \neq 0$. Again for a low density of uniform defects, we can choose large $d_x$, $d_y$ and $d_z$ values so that $\kappa_d \approx \kappa_m$ and $\alpha \cdot \left(\kappa_d - \kappa_m\right)/\kappa_m \approx 0$. Eq. (\ref{1 over kappa2}) can then be simplified as
\begin{equation}
\frac{1}{\kappa_{m+d}} \approx \frac{1 - \rho \cdot \left(\kappa_d - \kappa_m\right)/\kappa_m}{\kappa_m}
\label{1 over kappa2 simple1}
\end{equation}
Eq. (\ref{1 over kappa2 simple1}) is further written as
\begin{equation}
\kappa_{m+d} \approx \frac{\kappa_m}{1 - \rho \cdot \left(\kappa_d - \kappa_m\right)/\kappa_m} = \kappa_m - \left(\kappa_m - \kappa_d\right) \cdot \rho = \kappa_m \left(1-\eta \cdot \xi\right)
\label{kappa2 simple1}
\end{equation}
Eq. (\ref{kappa2 simple1}) is the same as Eq. (\ref{kappa2 simple}). 

\section{Combined Effects of Defects and Finite Length between Heat Source and Sink-surface}\label{length conversion}

In the axial defect density series of MD simulations, samples of different lengths $L$ were used to change the defect density. However, the resulting thermal conductivity is affected by both sample length \cite{M1992,PB1994,OS1999,ZJA2010a,ZJA2010b,ZAJGS2009} and defect density. Here we derive a thermal conductivity expression to separate the length and defect effects.

As shown in Fig. \ref{defect and end}, we assume that heat flows through a medium with a finite length $L$ that contains a defect in the middle. If we use a large $d_x$ to fully contain the defect effect and a large $\omega$ to contain the end surface effect, then the overall thermal conductivity of the defective section and the end surface regions can be taken as constants $\kappa_d$ and $\kappa_e$ respectively and the conductivity of the remaining material is that of the infinite bulk $\kappa_{m,\infty}$. Note that because both defect and end surface effects are considered here, we use $\kappa_{m,\infty}$ instead of $\kappa_{m}$ to emphasize that the thermal conductivity of the matrix is free of both defects and end surfaces (i.e., the length $L$ is infinite). Because heat flows through different regions in serial, the overall thermal resistivity is a length weighted average:
\begin{figure}
\includegraphics[width=6in]{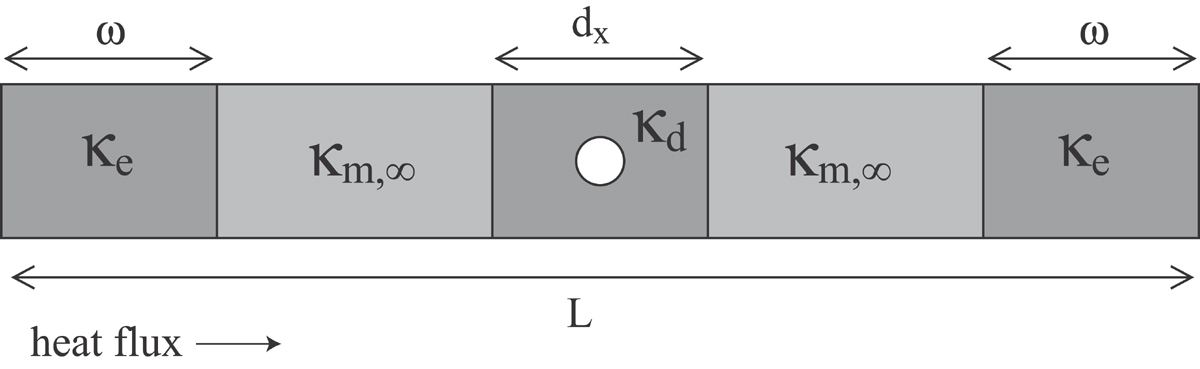}
\caption{Effects of defect and end surfaces on thermal transport.
\label{defect and end}}
\end{figure}
\begin{eqnarray}
\frac{1}{\kappa_{m+d}} &=& \frac{2\omega}{L} \cdot \frac{1}{\kappa_e} + \frac{d_x}{L} \cdot \frac{1}{\kappa_d} + \frac{L - 2\omega - d_x}{L} \cdot \frac{1}{\kappa_{m,\infty}} \nonumber \\ &=& \frac{2 \omega \left(\kappa_{m,\infty} - \kappa_{e}\right)}{L \cdot \kappa_{m,\infty} \cdot \kappa_{e}} + \frac{d_x \left(\kappa_{m,\infty} - \kappa_d \right)}{L \cdot \kappa_{m,\infty} \cdot \kappa_d} + \frac{1}{\kappa_{m,\infty}} \nonumber \\ &=& \frac{1}{\kappa_{m,\infty}} + \frac{p}{L} + q \cdot \rho = \frac{1}{\kappa_{m,\infty}} + \frac{p}{L} + q \cdot f^{-1} \cdot \xi 
\label{1 over defect kappa1}
\end{eqnarray}
where $p = 2 \omega \cdot \left(\kappa_{m,\infty} - \kappa_e\right) \cdot \kappa_{m,\infty}^{-1} \cdot \kappa_e^{-1}$, $q = \left(\kappa_{m,\infty} - \kappa_d\right) \cdot \kappa_{m,\infty}^{-1} \cdot \kappa_d^{-1}$ are constants for given types of surfaces (or interfaces) and defects, and $\rho$ is taken to be $\rho = d_x/L$ assuming $\alpha = 1$ (i.e., we can use the entire cross-section area to contain the areal scattering effects of defects). Eq. (\ref{1 over defect kappa1}) can be used to separate defect and system length effects.


\end{document}